\documentclass[
superscriptaddress,
preprint,
showpacs,
amsmath,
amssymb,
aps,
prb,
]{revtex4}

\usepackage{graphicx}% Include figure files
\usepackage{dcolumn}% Align table columns on decimal point
\usepackage{bm}% bold math
\usepackage{hyperref}% add hypertext capabilities
\begin{document}

%\preprint{}

\title{A New Method for Characterizing Bulk and Surface Conductivities of Three-Dimensional Topological Insulators: Inverted Resistance Measurements}

\author{Y. S. Eo}
	\email{eohyung@umich.edu}
	\affiliation{University of Michigan, Dept.~of Physics, Ann Arbor, Michigan 48109-1040, USA}
\author{K. Sun}
	\affiliation{University of Michigan, Dept.~of Physics, Ann Arbor, Michigan 48109-1040, USA}
\author{\c{C}. Kurdak}  
	\affiliation{University of Michigan, Dept.~of Physics, Ann Arbor, Michigan 48109-1040, USA}
\author{D.-J. Kim}
	\affiliation{University of California at Irvine, Dept.~of Physics and Astronomy, Irvine, California 92697, USA}
\author{Z. Fisk}
	\affiliation{University of California at Irvine, Dept.~of Physics and Astronomy, Irvine, California 92697, USA}
 
\date{\today}

\begin{abstract}

We introduce a new resistance measurement method that is useful in characterizing materials with both surface and bulk conduction, such as three-dimensional topological insulators. The transport geometry for this new resistance measurement configuration consists of one current lead as a closed loop that fully encloses the other current lead on the surface, and two voltage leads that are both placed outside the loop. We show that in the limit where the transport is dominated by the surface conductivity of the material, the four-terminal resistance measured from such a transport geometry is proportional to $\sigma_b/\sigma_s^2$, where $\sigma_b$ and $\sigma_s$ are the bulk and surface conductivities of the material, respectively. We call this new type of measurement \textit{inverted resistance measurement}, as the resistance scales inversely with the bulk resistivity. We discuss possible implementations of this new method by performing numerical calculations on different geometries and introduce strategies to extract the bulk and surface conductivities. We also demonstrate inverted resistance measurements on SmB$_6$, a topological Kondo insulator, using both single-sided and coaxially-aligned double-sided Corbino disk transport geometries. Using this new method, we are able to measure the bulk conductivity, even at low temperatures, where the bulk conduction is much smaller than the surface conduction in this material. 

\end{abstract}

\pacs{72.10.Bg, 71.27.+a, 73.20.-r}%

\maketitle

\section{\label{sec:Intro}Introduction}

 Electrical transport measurements are useful for studying new quantum materials, as they reveal critical information about low energy excitations of the quantum ground state. Characterizing transport is also a crucial step toward realizing these quantum materials in novel electronic devices. Typically, specific transport geometries are used for extracting valuable information from the material. For example, the Hall bar geometry is used for measuring the diagonal and off-diagonal (Hall) resistivity\cite{Hall1879}, and the Corbino disk geometry is used for measuring the diagonal conductivity\cite{Corbino1911,kleinman1960}. In the case of two-dimensional materials and thin conducting films, the diagonal and off-diagonal resistivity can also be extracted from arbitrarily shaped samples by using the van der Pauw method, which works well as long as the contacts at the edges are small enough\cite{VanPauw1958}. When multiple conducting channels are present in the sample, the measured conductivity is the sum of the conductivities of each channel. In some cases, the carrier density and mobility of each channel can even be individually determined using a method known as the mobility spectrum analysis\cite{Vurgaftman1998}.

 In the case of isotropic three-dimensional crystals, its conductivity can be found using four-terminal resistance measurements. The measured resistance is inversely proportional to the conductivity of the material with a prefactor that depends on the geometry of the sample and the position of the contacts. Except for special cases such as a long wire or a thin film, this geometric prefactor can only be determined by numerical calculations. Researchers, especially in the correlated electron community, sometimes perform transport measurements on raw crystals that have irregular shapes. Commonly they do not supplement the transport measurements with numerical calculations needed to determine the geometric prefactor because of the complicated details of the geometry. Instead, they report the bulk resistivity normalized to the room temperature resistivity, $\rho(T)/\rho(300 ~\mathrm{K})$, because this can be found easily by the measured resistance ratio $R(T)/R(300 ~\mathrm{K})$. In most cases, the material can be identified as either a conductor or an insulator by extrapolating $\rho(T)/\rho(300 ~\mathrm{K})$ to $T=0$ K. When the material under investigation is purely a bulk conductor, the reporting of the resistance ratio is a useful practice. On the other hand, as we will discuss below, resistance ratios are meaningless when the material has both surface and bulk conduction. 
 
 In the last decade, a new class of three-dimensional materials has been discovered; these are expected to have an insulating bulk and conducting surface states\cite{FuPRL}. These materials are called three-dimensional topological insulators (3D TIs)\cite{FuPRL,JEMoorePRB}. The surface conduction in these materials arises from the non-trivial topology of the bulk band, and its existence is robust. Electrical characterization of 3D TIs is challenging, as they are expected to have both bulk and surface conduction at finite temperatures. Resistance measurements from standard transport geometries do not provide information about the fraction of current flow through surface and bulk of the material. Furthermore, in situations where the surface conduction is significant, the bulk resistivity ratio, $\rho(T)/\rho(300 \mathrm{K})$, cannot be obtained from $R(T)/R(300 \mathrm{K})$ because the prefactor now also changes with temperature. We, therefore, recommend the physics community to stop the practice of using resistance ratios in characterizing such materials.
 
 For many of the weakly correlated 3D TIs, including Bi$_{1-x}$Sb$_{x}$\cite{DHsieh}, Bi$_2$Se$_3$\cite{YXiaNatPhys}, and Bi$_2$Te$_3$\cite{Chen2009}, bulk conduction by unintentional impurities is large enough to overwhelm the surface conduction\cite{Brahlek2015}. Because the electric current flows mostly through the bulk in this case, the surface conduction is difficult to explore. To extract the surface conductivity in these materials, one typically prepares ultrathin samples, since the effective conductivity from the measured resistance is normally $\sigma_{\mathrm{eff}}  \approx \sigma_s  + \sigma_{b} t$ ($\sigma_b$ is the bulk conductivity, $\sigma_s$ is the surface conductivity, and $t$ is the thickness of the sample). In principle, both $\sigma_b$ and $\sigma_s$ can be determined if one performs transport measurements on multiple samples with a wide range of thicknesses. 
 
 There are also materials in which, at low temperatures, the surface conduction dominates the transport. A good example is samarium hexaboride (SmB$_6$), which is under investigation as the first correlated 3D TI material\cite{DzeroPRL,Takimoto}.  The bulk gap of SmB$_6$ is clean and results from the hybridization between the $d$-electrons and the $f$-electrons at cryogenic temperatures\cite{MartinTheory1979,DzeroPRL}. The bulk conducts only through thermal excitation, and therefore the conductivity of the bulk decreases as the temperature decreases. As the temperature decreases further, the material undergoes a bulk-to-surface crossover, and the current begins to flow more on the surface than the bulk because the bulk conductivity is too small\cite{JWAllen1979,WolgastPRB}. Below this crossover temperature, the bulk conductivity is challenging to obtain because the measured resistance is mostly governed by the surface.
 
 In this paper, we introduce a generic transport strategy, which we call the \textit{inverted resistance measurement}, to find the bulk conductivity contribution when the surface conduction dominates the bulk. Next, we analyze specific transport geometries that are suitable for the inverted resistance measurements. Finally, we demonstrate this new method experimentally on single crystals of SmB$_6$ and show that the bulk conductivity can be extracted below a crossover temperature where conduction is dominated by surface states. 

\section{\label{sec:Formalism}Formalism for Resistance}
	
In this section, we introduce a general formalism for understanding resistance measurements in an isotropic 3D material with both bulk and surface conduction, such as a 3D topological insulator. In this formalism, any four-terminal resistance can be expressed in a suitable dimensionless function that depends on the ratio of the bulk and surface conductivities. We further consider this dimensionless function in two extreme regimes, $\sigma_{b}l \gg \sigma_{s} $ and $\sigma_{b}l \ll \sigma_{s} $, using a series expansion, where $l$ is the characteristic length parameter. We will propose a generic transport geometry with a four-terminal resistance configuration that allows us to access the bulk conductivity even while the surface dominates the conduction. 

\subsection{\label{subsec:PerturbApproach}Perturbative Approach of Scalable Resistance}
In general, any resistance measurement, say $R$, of a 3D material is a function of the bulk ($\sigma_{b}$) and surface conductivity ($\sigma_{s}$) for a given transport geometry, i.e., $R = R(\sigma_{b},\sigma_{s})$. The bulk and surface conductivity have different units of $1/(\mathrm{\Omega}\cdot \mathrm{m})$ and $1/\mathrm{\Omega}$, respectively. The resistance can be rewritten in the following form:
\begin{equation}
	R=\frac{1}{\sigma_{s}}f(x) \quad \textrm{and} \quad x=\frac{\sigma_{b}l}{\sigma_{s}},
	\label{Eq:ScaleResistance}
\end{equation}
where $f(x)$ is a dimensionless function that depends on the current distribution for the transport geometry, and $l$ is some characteristic length that depends on the transport geometry. The parameter, $x$, is also dimensionless, and it is defined by the ratio between  $\sigma_{b}$ and  $\sigma_{s}$ multiplied by $l$. Resistance is determined by the spatial dependence of the current or electrical potential of the transport geometry. This spatial dependence is uniquely determined by the boundaries, i.e., bulk/surface interface, and therefore the ratio between the surface and bulk conductivity determines $f(x)$. When $x$ remains constant, $R$ scales with either $1/\sigma_{b}$  or  $1/\sigma_{s}$. In Eq.~(\ref{Eq:ScaleResistance}), we choose $1/\sigma_{s}$ so that the function $f(x)$ becomes a dimensionless function.

We have now expressed the resistance in a form that depends on $\sigma_{s}$ and $x$, instead of on $\sigma_{b}$ and $\sigma_{s}$. Expressing the resistance in the form of Eq.~(\ref{Eq:ScaleResistance}) is powerful when we consider a 3D topological insulator in two extremes: the bulk-dominated regime ($\sigma_{b}l \gg \sigma_{s} $) and the surface-dominated regime ($\sigma_{b}l \ll \sigma_{s} $). Both regimes can also be considered in the asymptotic limits of the dimensionless function, $f(x)$: the bulk-dominated regime can be studied in the $x\rightarrow\infty$ limit, and the surface-dominated regime can be studied in the $x\rightarrow0$ limit.

In the bulk-dominated regime, the case when current flows mostly in the bulk, we can expand the function $f(x)$ in powers of $1/x$, $f(x)=C_{-1}(1/x)+C_{-2}(1/x)^2+\dotsb $, where $C_{-1}$, $C_{-2}$, $\dotsc$ are coefficients that depend on the geometry of the transport. The resistance is therefore:
\begin{equation}
	R=\frac{C_{-1}}{\sigma_{b}l}+\frac{C_{-2}}{(\sigma_{b}l)^2}\sigma_{s}+\dotsb.
	\label{Eq:BulkDomSeries}
\end{equation}
The first order term in Eq.~(\ref{Eq:BulkDomSeries}) only depends on the bulk properties, and this term usually overwhelms the higher order terms in resistance measurements of conventional transport geometries. The higher order terms, which contain $\sigma_{s}$, are therefore difficult to measure. 

In the surface-dominated regime, the case when the current flows mostly on the surface, we can use the following asymptotic form $f(x)=C_{0}+C_{1}x+C_{2}x^2+\dotsb $, where $C_{0}$, $C_{1}$, $C_{2}$, $\dotsc$ are coefficients. Thus, the resistance is:
\begin{equation}
	R=\frac{C_{0}}{\sigma_{s}}+\frac{C_{1}}{\sigma_{s}^2}\sigma_{b}l+\dotsb.
	\label{Eq:SurfDomSeries}
\end{equation}
In Eq.~(\ref{Eq:SurfDomSeries}), the first order term depends only on surface properties, and this term usually overwhelms the higher order terms in conventional transport measurements. Note that the bulk conductivity only arises in higher order terms. Although measuring the higher order terms is desirable for accessing the bulk conductivity, this is usually not possible in conventional transport measurements since the first term dominates.

Note that both asymptotic equations, Eq.~(\ref{Eq:BulkDomSeries}) and Eq.~(\ref{Eq:SurfDomSeries}), fail to cover the range near $x\approx1$, which is the bulk-to-surface crossover regime, where the bulk and surface conduction are comparable. However, if the temperature dependence of $\sigma_{s}$ is weak compared to $\sigma_{b}$, we can make use of the standard two-channel model that experimentalists conventionally use to cover this range. When the sample is sufficiently thin, the following relation holds for most conventional resistance geometries: $R\propto 1/(\sigma_{s}+\sigma_{b}l)$. This relation is useful in connecting the bulk and surface dominated regimes when extracting the bulk conductivity. 

\subsection{\label{subsec:Inverted}Inverted Resistance Measurement}

\begin{figure}[t]
\begin{center}
\includegraphics{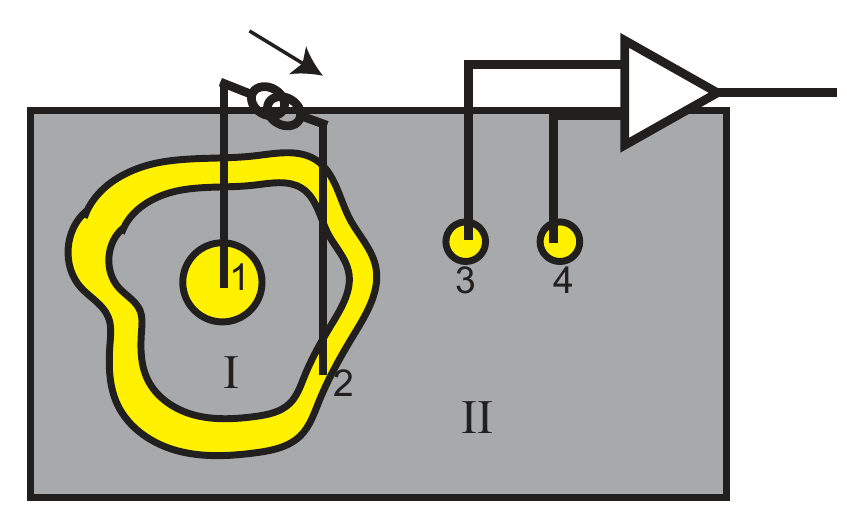}
\caption[Schematic diagram of a generic inverted resistance measurement.]{Schematic diagram of a generic inverted resistance measurement. In such a measurement, $R_{1,2;3,4}$, current lead 2 must fully enclose lead 1, and the voltage leads must be placed outside of the loop defined by lead 2.}
\label{Fig:GenericInverted}
\end{center}
\end{figure}

In this subsection, we introduce a new non-local transport measurement that is extremely powerful for characterizing a 3D TI with a small bulk conductivity. The bulk conductivity can be extracted in the surface-dominated regime, $\sigma_{b}t\ll \sigma_{s}$ (for convenience we choose $l=t$,  $t$ is thickness), by totally suppressing the first order term (surface term) in Eq.~(\ref{Eq:SurfDomSeries}). 

This new transport measurement configuration consists of two current leads, one fully enclosing the other as shown in Fig.~(\ref{Fig:GenericInverted}). Since the current contact 2 forms a closed loop, it separates the surface into two regions: I and II. Terminal 1 is in region I, terminal 2 is on the loop, and the other two terminals are in region II. We consider a resistance measurement in this transport geometry as the following: while passing current between terminal 1 and 2, the voltage is measured between terminals 3 and 4 (i.e., $R_{1,2;3,4}  = V_{3,4}  / I_{1,2}$). This defines the inverted resistance measurement.

Let us first consider what happens if we employ this geometry to characterize a two-dimensional electron gas or a thin film. When we connect leads 1 and 2 to the current source as shown in Fig.~(\ref{Fig:GenericInverted}), current will flow only in region I. Here, the metallic loop (contact 2) would act like a two-dimensional Faraday cage blocking all the electric field from inside and thus there would be no current flow or electric field in region II. If the voltage leads are placed in region II, then $V_{3,4}  = 0$. Because of this, such an inverted resistance measurement is not useful as $R_{1,2;3,4}$, would always be zero, regardless of the surface conductivity, and thus cannot be used to characterize a two-dimensional electron gas or thin film.

Because lead 2 acts as a Faraday cage only for the surface, if one employs this inverted resistance measurement on a bulk conductor, current will flow everywhere in the sample. Thus, one would expect a small but a measurable $V_{3,4}$. However, this geometry would not be used to measure bulk conductivity, since it would require numerical calculations of fringe fields and would not offer any benefits over conventional transport measurements.

On the other hand, in the case of a 3D TI, where one needs to characterize both surface and bulk conductivities, this type of inverted resistance measurement can provide information that cannot be accessed by conventional resistance measurements. To illustrate the power of this method when applied to a 3D TI, let us consider this measurement using the formalism introduced in the previous subsection. If the bulk conductivity is sufficiently low that the 3D TI is in the surface-dominated regime ($\sigma_{b}l \ll \sigma_{s}$), even for a sample with finite thickness, $R_{1,2;3,4}$ suppresses the leading order (surface term) term of Eq.~(\ref{Eq:SurfDomSeries}), and therefore the second term dominates ($R\propto \sigma_{b}t/\sigma_{s}^2$). Notice that the second order term in Eq.~(\ref{Eq:SurfDomSeries}) is proportional to the bulk conductivity.

To justify why the first term vanishes, let us first consider the case when $\sigma_{b}=0$. In $\sigma_{b}=0$, every term vanishes except the first term in Eq.~(\ref{Eq:SurfDomSeries}). In Fig.~(\ref{Fig:GenericInverted}), the loop (terminal 2) must capture the entire current from terminal 1, since the surface is the only current path available. Inside region I, the electric potential drop is proportional to  $1/\sigma_{s}$, whereas the entire region II must be equipotential to terminal 2 ($V$ = 0 when grounded). Then, $V_{3,4} = 0$. Therefore, we find that $C_{0}  = 0$ in Eq.~(\ref{Eq:SurfDomSeries}). When $\sigma_{b}\neq0$, this resistance measurement, $R_{\mathrm{Inv}}$ (inverted resistance) can be written as: 
\begin{equation}
	R_{\mathrm{Inv}}=0+\frac{C_{1}}{\sigma_{s}^2}\sigma_{b}t+\dotsb.
	\label{Eq:InvSeries}
\end{equation}
Thus, we conclude that whenever a transport geometry on a 3D TI utilizes a closed loop on the surface and the voltage is measured outside of that loop, then the highest order term (surface term) in Eq.~(\ref{Eq:SurfDomSeries}) is suppressed and the next leading order term, which contains the bulk conductivity, dominates. Therefore, the inverted resistance measurement can be used to access the bulk conductivity in situations where the surface conduction dominates the bulk. 

In this subsection, we have introduced the inverted resistance measurement, which provides the bulk conductivity information even in the presence of strong surface conduction. When we combine this inverted resistance measurement with conventional transport experiments, we can extract bulk and surface conductivities. In the following section (Sec.~\ref{sec:TransportGeometry}), we introduce specific sample geometries and present the geometric factors that are found by numerical calculations. We also introduce strategies to implement the inverted resistance and the numerical results in experiment to extract the conductivities. 

\section{\label{sec:TransportGeometry}Transport Geometries for Inverted Resistance Measurements}

\begin{figure}[t]
\begin{center}
\includegraphics{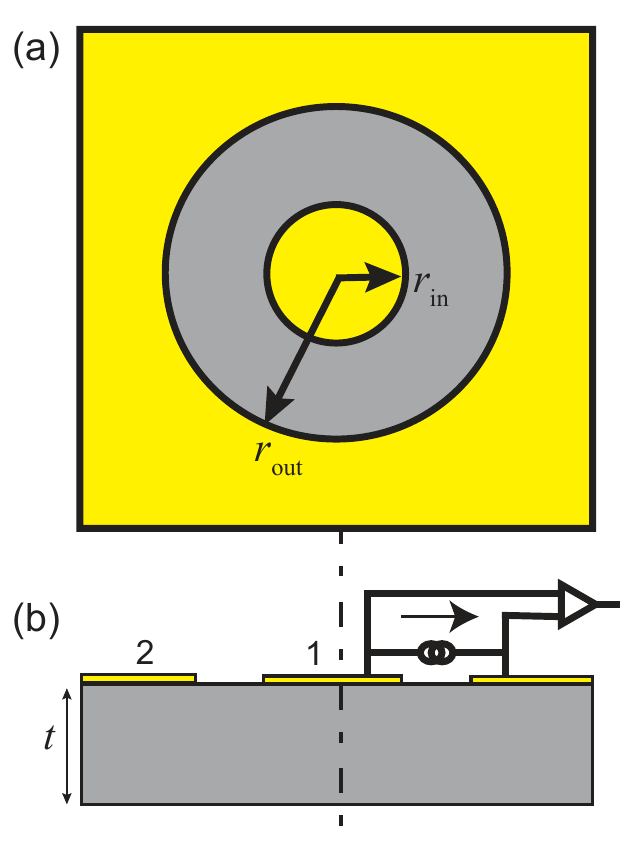}
\caption[Standard two-terminal Corbino disk.]{A standard two-terminal Corbino disk on a sample. (a) Top view: the sample is shown in gray and the highly conductive contacts are shown in yellow. (b) Side view: the current and voltage amplifier that is needed to perform the resistance measurements are connected to terminals 1 and 2.}
\label{Fig:StandTwoTermCorbino}
\end{center}
\end{figure}

\begin{figure}[t]
\begin{center}
\includegraphics{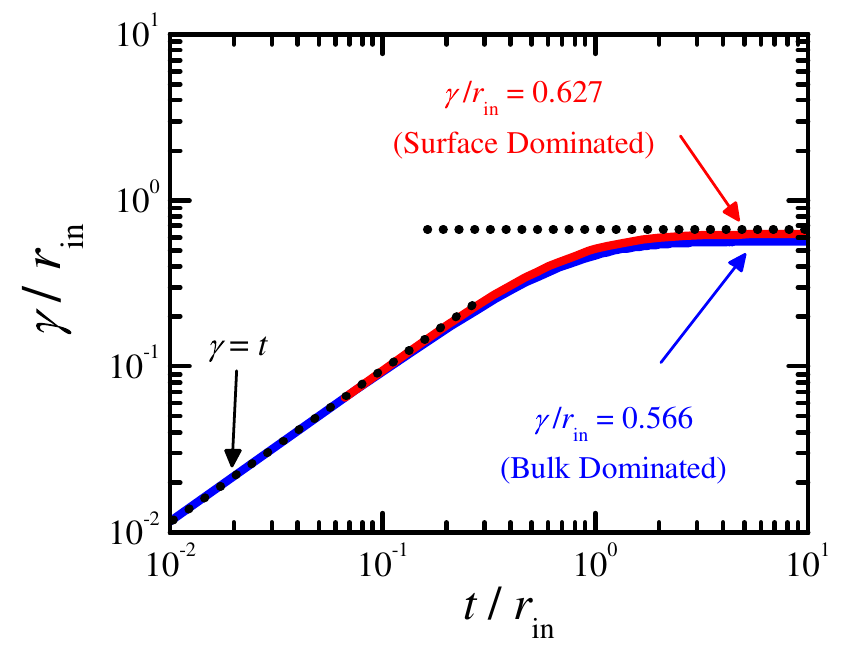}
\caption[Effective thickness vs$.$ actual thickness in a standard Corbino disk]{Effective thickness, $\gamma$, in the two-channel model vs$.$ real sample thickness, $t$, for a standard Corbino disk. Both axes are divided by the inner radius ($r_{\mathrm{in}}=150$ $\mathrm{\mu m}$). The solid red line is the numerical calculation in the surface-dominated regime when only the top surface contributes. When the thickness is very large, $\gamma/r_{\mathrm{in}}$ asymptotically approaches 0.627. The solid blue line is the numerical calculation in the bulk-dominated regime. At large thickness, $\gamma/r_{\mathrm{in}}$ asymptotically approaches 0.566. The effective thickness approaches to the true thickness of the sample when the sample thickness is sufficiently thin ($\gamma = t$).}
\label{Fig:StandTwoTermCorbinoResult}
\end{center}
\end{figure}

Before we introduce transport geometries that allow inverted resistance measurements, we review a simple geometry, the Corbino disk, which employs a closed circular current loop shown in Fig~(\ref{Fig:StandTwoTermCorbino})~(a). To remind the reader of the standard Corbino disk measurement, the current flows from the center to the outer loop radially, and the voltages are measured either at the two current terminals (two-terminal resistance) or at two points within the transport region whose radii are $r_{\mathrm{in}}$~(inner radius) and $r_{\mathrm{out}}$~(outer radius) (four-terminal resistance). The two-terminal Corbino resistance measurement configuration is shown in Fig~(\ref{Fig:StandTwoTermCorbino})~(b). In a perfect 2D transport case, where current only flows on the surface ($\sigma_{b}=0$ and $\sigma_{s}\neq0$), the functional form of this standard resistance is well known: 
\begin{equation}
	R_{\mathrm{Corbino}}=\frac{1}{2\pi}\ln(\frac{r_{\mathrm{out}}}{r_{\mathrm{in}}})\frac{1}{\sigma_{s}}.
	\label{Eq:SurfCorbinoFormula4}
\end{equation}
In the presence of both bulk and surface conduction ($\sigma_{b}\neq0$ and $\sigma_{s}\neq0$), the two-channel model is a good approximation for the standard Corbino resistance: 
\begin{equation}
	R_{\mathrm{Corbino}}\approx\frac{1}{2\pi}\ln(\frac{r_{\mathrm{out}}}{r_{\mathrm{in}}})\frac{1}{\sigma_{s}+\sigma_{b}\gamma},
	\label{Eq:CorbinowithBulk}
\end{equation}
where $\gamma$ is the effective thickness of the sample. $\gamma$ depends on how the current flows in the defined geometry. We study numerically how the effective thickness changes as function of true sample thickness by performing finite element analysis calculations using Comsol Multiphysics AC/DC module. The results from the bulk-dominated regime are shown in a solid blue line in Fig~(\ref{Fig:StandTwoTermCorbinoResult}). In the thin sample limit, where the thickness of the sample is much smaller than both the inner radius and the annular region (surface transport region) of the Corbino disk ($r_{\mathrm{in}}\gg t$ and $r_{\mathrm{out}}-r_{\mathrm{in}}\gg t$),  $\gamma$ approximates to the true thickness of the sample, $t$. In the very thick limit ($r_{\mathrm{in}}\ll t$ and $r_{\mathrm{out}}-r_{\mathrm{in}}\ll t$), $\gamma$ is independent of $t$. 

It is important to note that for a given sample, the effective thickness can be slightly different in the bulk- or surface-dominated transport regimes. To illustrate this, we have performed a series of numerical calculations by adding a surface channel with a broad range of conductivity on the top surface of the sample. The result is shown in a solid red line, which is similar, but not identical, to the results of the bulk-dominated regime. In the very thick limit, the effective thickness is again independent of $t$, but with a value that is about 10$\%$ larger than that of the bulk-dominated regime. We were also able to analyze a Corbino disk geometry on an infinite thickness sample using analytical methods. We find the asympototic value of effective thickness extracted from the numerical calculations agrees with the analytical study in the infinite thickness limit; the analytical derivation is presented in Appendix~\ref{Appedix:AnalCorbino}. In the following subsections, continuing to use finite element analysis numerical calculations, we will consider different extensions of Corbino disk geometries that are suitable for the inverted resistance measurement.

\subsection{\label{subsec:SingleFourTerminalInv}Single-Sided Four-Terminal Corbino disk}

\begin{figure}[t]
\begin{center}
\includegraphics{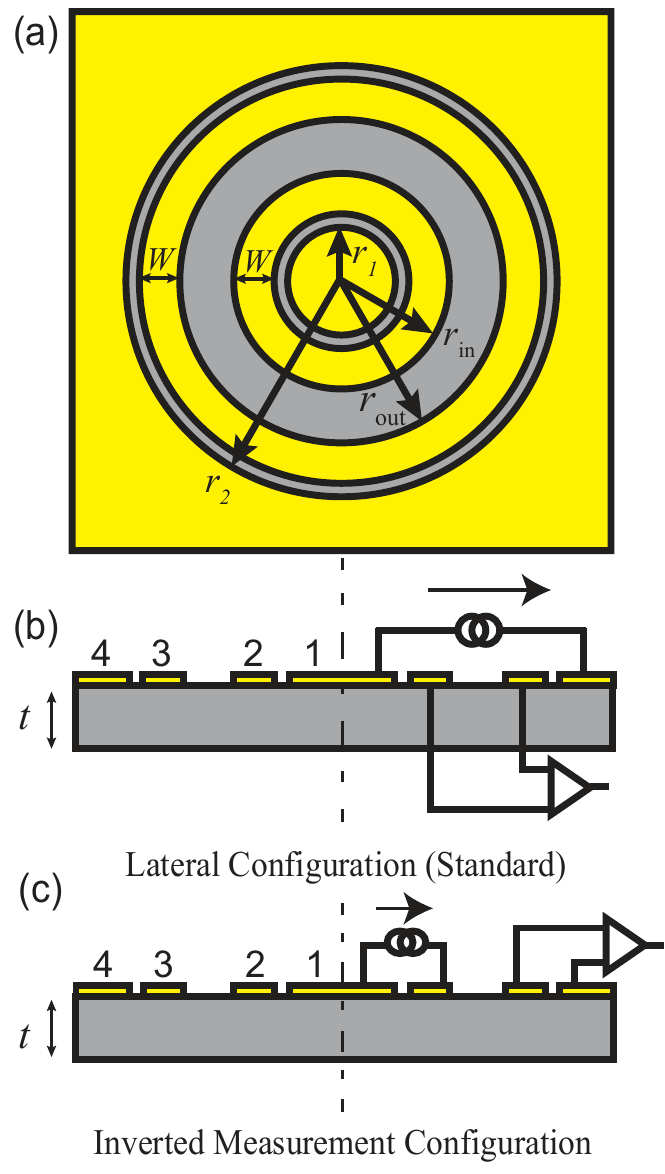}
\caption[Four-terminal single-sided Corbino disk.]{Four-terminal single-sided Corbino disk. The sample is shown in gray and the highly conductive contacts are shown in yellow. (a) Top view. (b) Side view and resistance configuration of the lateral configuration. (c) Side view and the resistance configuration of the inverted measurement configuration. We choose the dimensions: $r_{1}$ = 100 $\mathrm{\mu m}$, $r_{2}$ = 800 $\mathrm{\mu m}$, $r_{\mathrm{in}}$ = 200 $\mathrm{\mu m}$, $r_{\mathrm{out}}$ = 300 $\mathrm{\mu m}$, and $W$ = 75 $\mathrm{\mu m}$.}
\label{Fig:FourTermCorbino}
\end{center}
\end{figure}

In this subsection, we consider a transport geometry consisting of a Corbino disk with two metallic rings in the annular region, as shown in Fig.~(\ref{Fig:FourTermCorbino}). This transport geometry can be realized by a single step of lithography. The top view of the sample, the surface where the Corbino disk is patterned, is shown in Fig.~(\ref{Fig:FourTermCorbino})~(a). On top of the sample surface (shown in grey), the transport geometry pattern is defined by highly conductive metal contacts such as gold (shown in yellow). In addition to the inner- (terminal 1) and outer- (terminal 4) circular metallic regions, there are two metallic rings (terminals 2 and 3), each with width, $W$. The side views of two different measurement configurations are shown in Fig.~(\ref{Fig:FourTermCorbino})~(b) and Fig.~(\ref{Fig:FourTermCorbino})~(c). In the lateral configuration shown in Fig.~(\ref{Fig:FourTermCorbino})~(b), the resistance is measured by passing current between terminals 1 and 4 and measuring the voltage between terminals 2 and 3, i.e., $R_{L}=V_{2,3}/I_{1,4}$. This measurement is equivalent to the standard measurement of a conventional Corbino disk. In the inverted configuration shown in Fig.~(\ref{Fig:FourTermCorbino})~(c), the resistance is measured by passing current between terminals 1 and 2 and measuring the voltage between terminals 3 and 4, i.e., $R_{\mathrm{Inv}}=V_{3,4}/I_{1,2}.$ Recall that $C_{0} = 0$ in Eq.~(\ref{Eq:InvSeries}) in this inverted resistance measurement, and therefore $R_{\mathrm{Inv}}=(C_{1}t/\sigma_{s}^2) \sigma_{b}$. 
\begin{figure}[p]
\begin{center}
\includegraphics{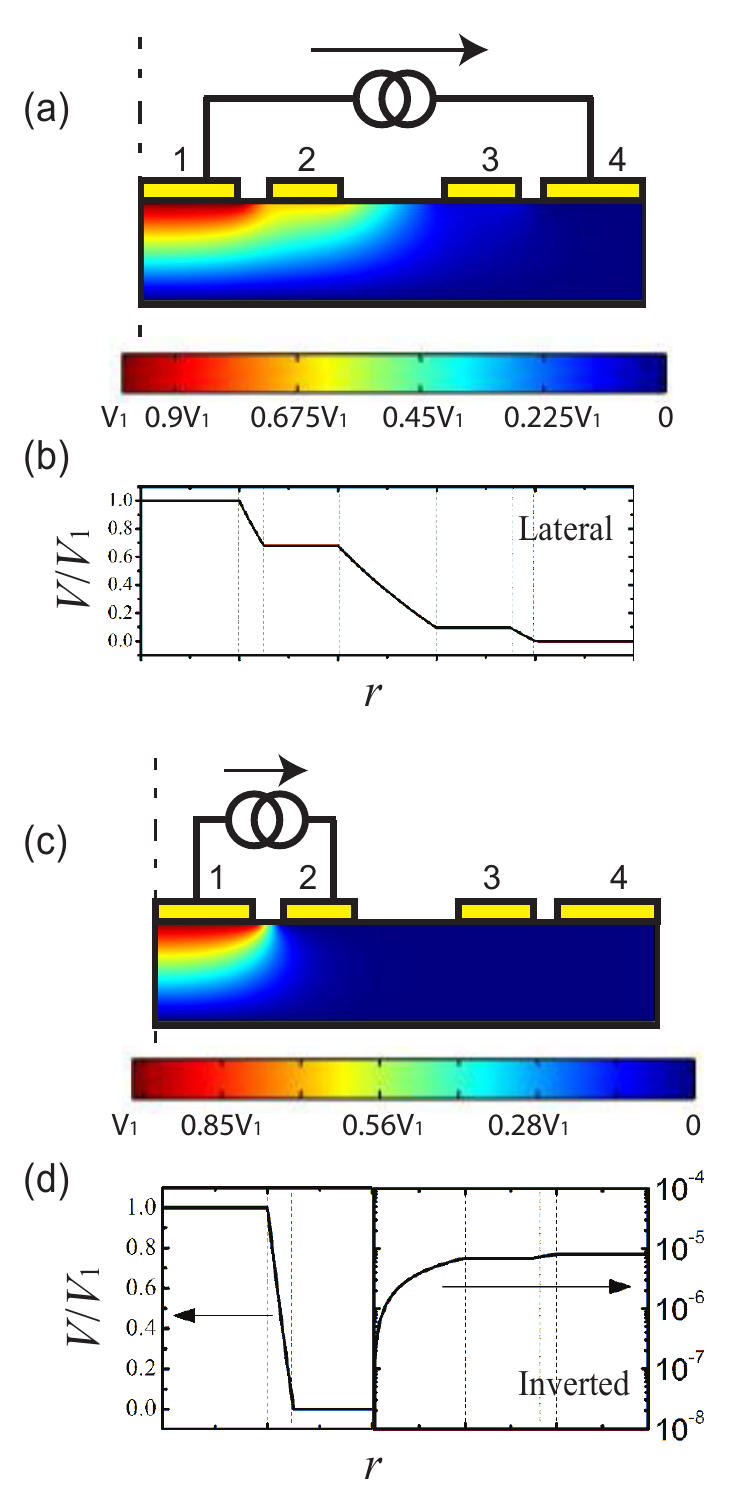}
\caption[The potential distributions of the single sided four-terminal Corbino disk.]{The potential distributions of the single-sided four-terminal Corbino disk. The bulk and on the surface is calculated in the surface-dominated transport regime ($\sigma_{s}\gg\sigma_{b}t$); $t$ = 100 $\mathrm{\mu m}$, $\sigma_{b}$=0.0013 1/($\mathrm{\Omega}\cdot$m), and $\sigma_{s}$ = 0.005 1/$\mathrm{\Omega}$. The equipotential values are normalized by the potential at the current source ($V_{1}$). (a) Potential distribution in the bulk when the current is connected to the lateral configuration. (b) Potential distribution on the surface as a function of radial position when the current is connected to the lateral configuration. (c) Potential distribution in the bulk when the current is connected to the inverted configuration. (d) Potential distribution on the surface as a function of radial position when the current is connected to the inverted configuration. The surface region in $r< r_{\mathrm{in}}$ is plotted in a linear scale, and $r> r_{\mathrm{in}}$ is plotted in a logarithmic scale.}
\label{Fig:PotDistribution}
\end{center}
\end{figure} 
 
When the current flows only on the surface, the lateral resistance, $R_{L}$, is identical to Eq.~(\ref{Eq:SurfCorbinoFormula4}). We rely on the numerical studies, using a finite element analysis software (Comsol Multiphysics AC/DC module), for the case when the current also flows in the bulk ($\sigma_{b}\neq0$ and $\sigma_{s}\neq0$). The numerical calculations are performed for a 100 $\mathrm{\mu}$m thick sample for both resistance configurations ($R_{L}$ and $R_{\mathrm{Inv}}$). The solution of electrical potential resulting from the current flow in the lateral configuration is shown for the bulk and surface in Fig.~(\ref{Fig:PotDistribution})~(a) and Fig.~(\ref{Fig:PotDistribution})~(b), respectively. In Fig.~(\ref{Fig:PotDistribution})~(a), the equipotential lines are coded in color. The bulk current flows normal to those equipotential lines. In Fig.~(\ref{Fig:PotDistribution})~(b), on the surface, the potential drops from the source to the ground on the surface logarithmically, except for in the metallic ring region, where the potential remains constant. 

 For the inverted resistance measurement, the electrical potential that is calculated numerically is shown for the bulk and surface in Fig.~(\ref{Fig:PotDistribution})~(c) and Fig.~(\ref{Fig:PotDistribution})~(d), respectively. The inverted resistance is expected to be much smaller than the lateral resistance as it results from the fringe field created near the outer part of the enclosed loop. The fringe fields are difficult to visualize in Fig.~(\ref{Fig:PotDistribution})~(c). However, the effect from the fringes can be seen when the potential on the surface is plotted on a logarithmic scale, as shown in Fig.~(\ref{Fig:PotDistribution})~(d). The plot indicates that the potential is gradually increasing as the  distance from the terminal 2 ring increases. 

\begin{figure}[t]
\begin{center}
\includegraphics{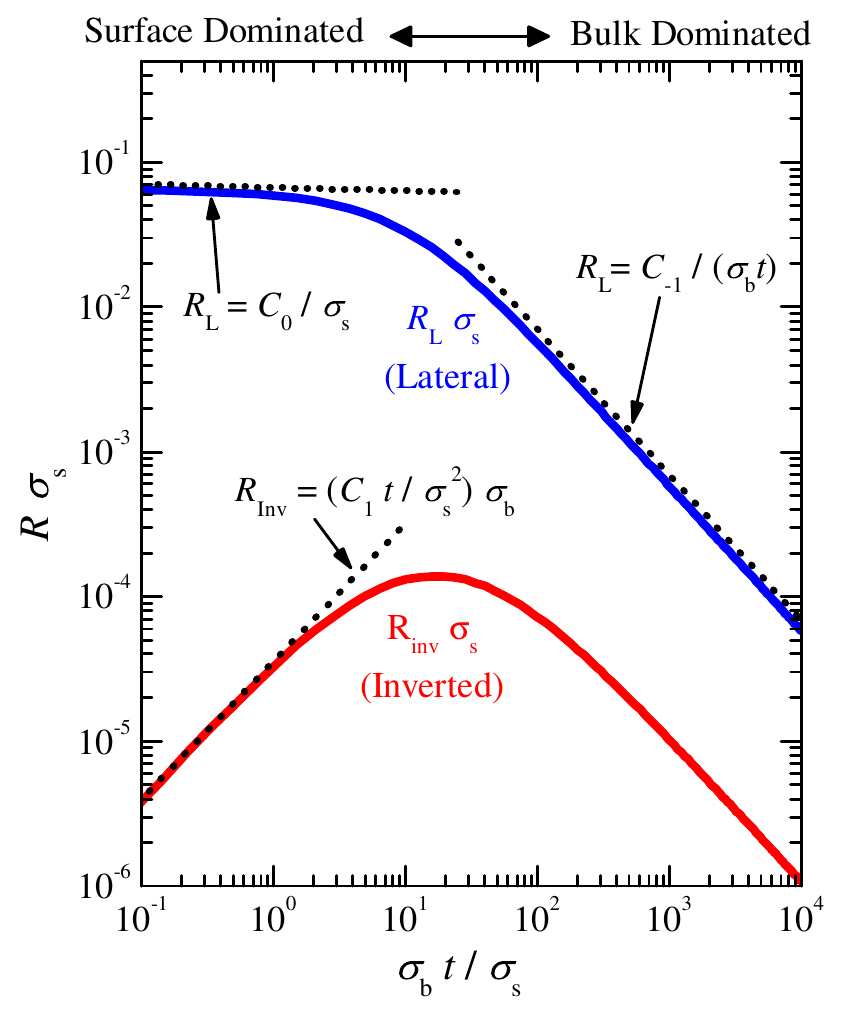}
\caption[Example of numerical results for a single-sided Corbino disk.]{Example of numerical results of a single-sided Corbino disk. $R_{L}\sigma_{s}$ (blue) and $R_{\mathrm{Inv}}\sigma_{s}$ (red) as a function of $x$ ($=\sigma_{b}t/\sigma_{s}$) when $t$ = 1000 $\mathrm{\mu}\mathrm{m}$. The dotted lines represent the asymptotic values from the first order term from Eq.~(\ref{Eq:BulkDomSeries}), Eq.~(\ref{Eq:SurfDomSeries}), and Eq.~(\ref{Eq:InvSeries}).}
\label{Fig:FourTermSideExample}
\end{center}
\end{figure}

We have performed these numerical calculations for different ratios of bulk to surface conductivity. In Fig.~(\ref{Fig:FourTermSideExample}), we plot the dimensionless function, $f(x)= R\sigma_{s}$, for both the inverted ($R_{\mathrm{Inv}}\sigma_{s}$) and the lateral resistance ($R_{L}\sigma_{s}$) as a function of $x$ ($=\sigma_{b}t/\sigma_{s}$). Most importantly, we find that $R_{L}\sigma_{s}\propto 1/x$ when $x\rightarrow\infty$, and $R_{\mathrm{Inv}}\sigma_{s} \propto x$  when $x\rightarrow0$. Therefore, in those two limits, the leading order terms in Eq.~(\ref{Eq:BulkDomSeries}) and Eq.~(\ref{Eq:InvSeries}) dominate. Furthermore, the flat line in $R_{L}\sigma_{s}$ when $x\rightarrow0$ agrees with Eq.~(\ref{Eq:SurfCorbinoFormula4}) and the leading order term of Eq.~(\ref{Eq:SurfDomSeries}) ($C_{0}=0.0645$ and $C_{0}\gg C_{1},C_{2},\dotsc$). 

\begin{figure}[t]
\begin{center}
\includegraphics{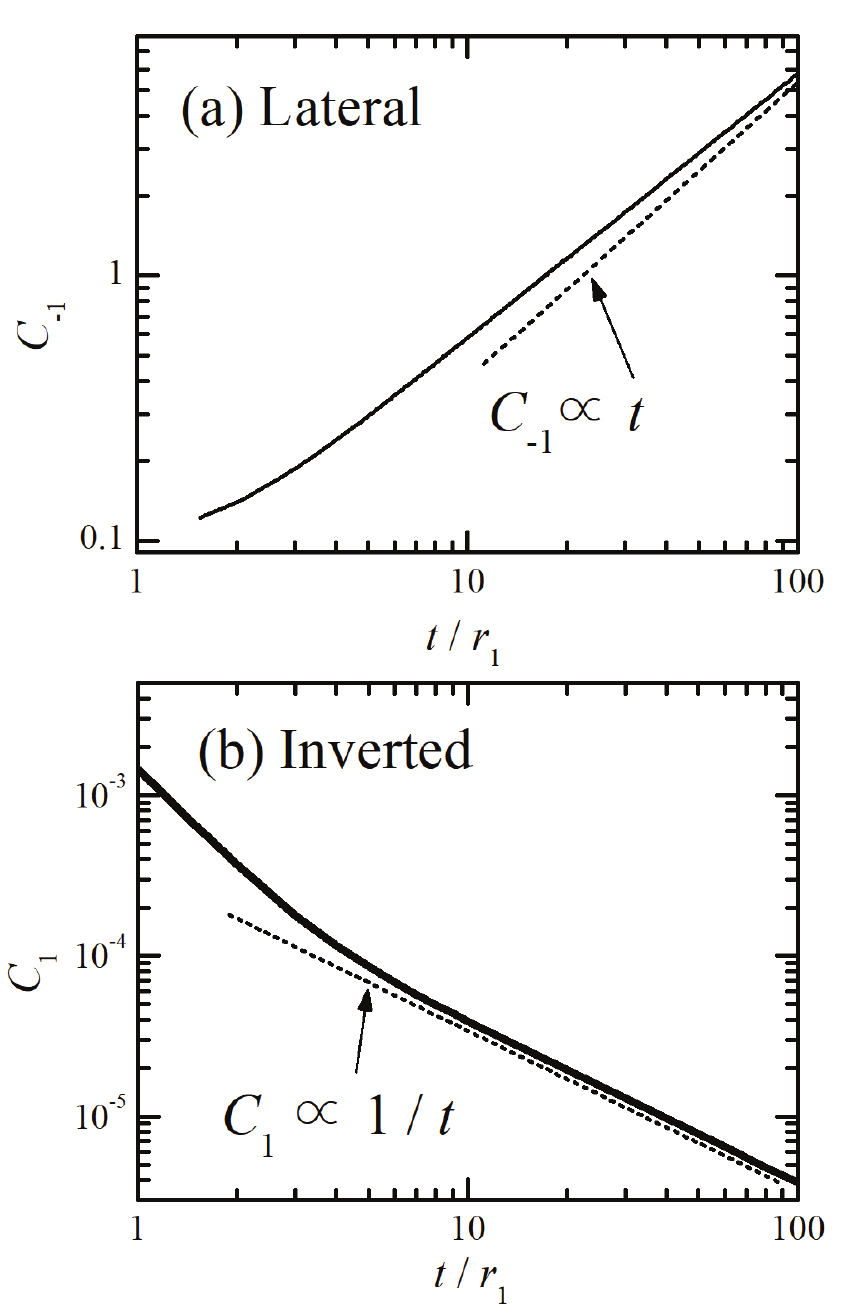}
\caption[The transport coefficients for the single-sided four-terminal Corbino disk geometry as a function of sample thickness.]{The transport coefficients for the single-sided four-terminal Corbino disk geometry as a function of sample thickness. (a) $C_{-1}$ of the lateral resistance $R_{L}$ in the bulk-dominated regime ($\sigma_{b}t\gg\sigma_{s}$). (b) $C_{1}$ of the inverted resistance $R_{\mathrm{Inv}}$ in the surface-dominated regime ($\sigma_{s}\gg\sigma_{b}t$). The thickness is dividing by $r_{1}$ (= 100 $\mathrm{•\mu m}$). The dotted lines indicate the transport coefficients when they are independent of thickness.}
\label{Fig:TransCoeffSingleCorbino}
\end{center}
\end{figure}

By iterating the numerical calculations for different thicknesses, the transport coefficients, $C_{-1}$ and $C_{1}$, are found as a function of thickness. Fig.~(\ref{Fig:TransCoeffSingleCorbino})~(a) shows the coefficient, $C_{-1}$, for the lateral measurement ($R_{L}$) in the bulk-dominated regime ($\sigma_{b}t\gg\sigma_{s}$) plotted as a function of dimensionless thickness ($t/r_{1}$). Fig.~(\ref{Fig:TransCoeffSingleCorbino})~(b) shows the coefficient, $C_{1}$, for the inverted measurement ($R_{\mathrm{Inv}}$) in the surface-dominated regime ($\sigma_{b}t\ll\sigma_{s}$). Note that at very large thicknesses, $C_{1}\propto1/t$, and $C_{-1}\propto t$, which means the resistances become independent of thickness. In the following section (Sec.~\ref{sec:Realization}), we will demonstrate this transport geometry on a SmB$_{6}$ sample, and use the results in Fig.~(\ref{Fig:TransCoeffSingleCorbino}) to find the bulk conductivity. In the following subsection, we consider a more advanced transport geometry design that results in a larger $C_{1}$ value and better confines the transport region. 

\subsection{\label{subsection:DoubleSidedInvert}Double-Sided Two-Terminal Corbino disks}

\begin{figure}[p]
\begin{center}
\includegraphics{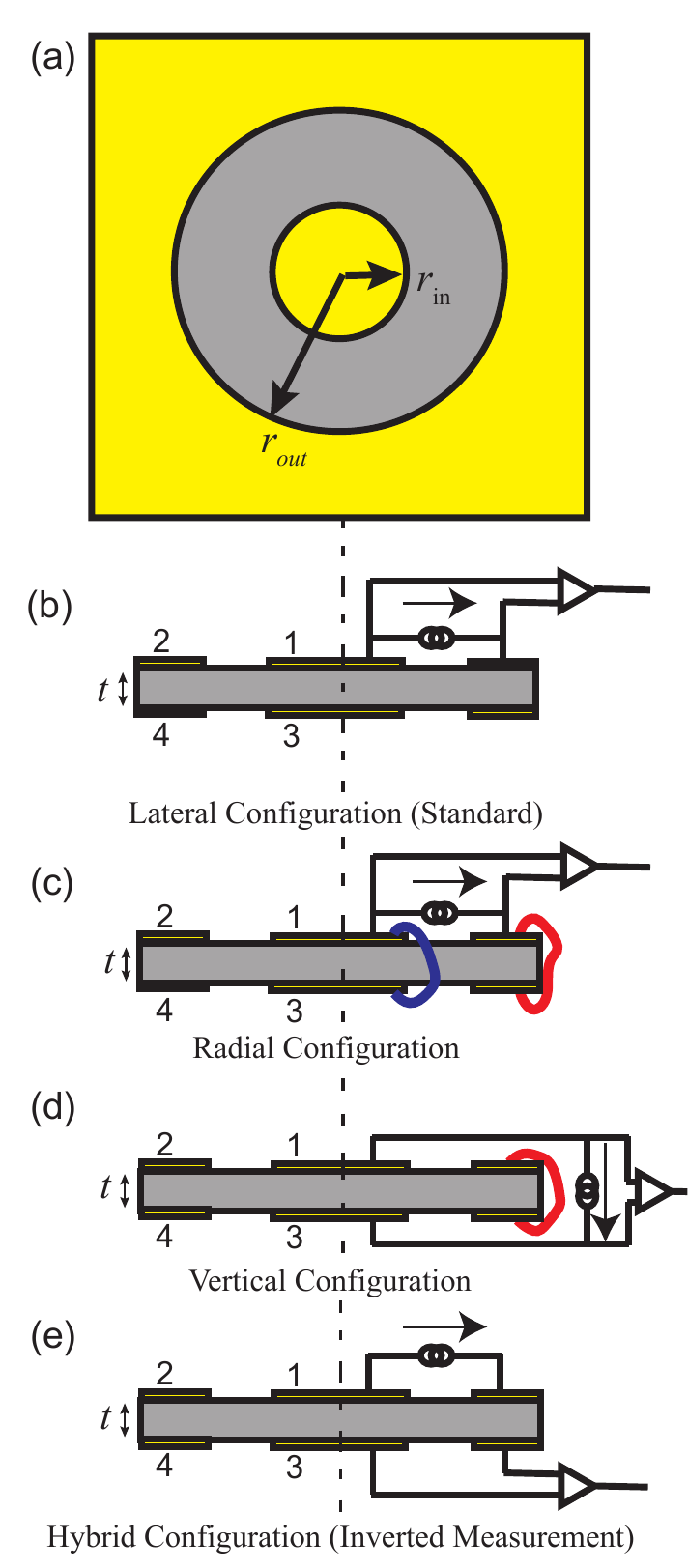}
\caption[Double-sided two-terminal Corbino disk]{Double-sided two-terminal Corbino disk. The sample is shown in gray and the highly conductive contacts are shown in yellow. (a) Top and bottom surface view. (b) Side view and the lateral resistance configuration. (c) Side view and the radial resistance configuration. The blue and red lines are jumper wires. (d) Side view and the vertical resistance configuration. (e) Side view and the hybrid resistance configuration. The hybrid measurement is an inverted resistance measurement in the surface-dominated regime ($\sigma_{s}\gg\sigma_{b}t$). We choose the Corbino dimensions as $r_{1}$ = 150 $\mathrm{\mu m}$ and  $r_{2}$ = 300 $\mathrm{\mu m}$.}
\label{Fig:DoubleSidedTwoTerminalCorbino}
\end{center}
\end{figure}

 In the previous subsection, we have considered a transport geometry defined on a single surface, and showed that the inverted resistance originates from the small fringe effects created near the enclosed loop. In this subsection, we consider a more advanced transport geometry, where we employ two coaxially aligned Corbino disks on opposite surfaces. We show that in this case, there is an inverted resistance measurement configuration that measures the current reaching the opposite surface, and this contribution can be much larger than the fringe effects on a single surface. Furthermore, since there are two Corbino disks placed on opposite surfaces, this geometry would allow us to measure the conductivity of both surfaces. This can also be a powerful geometry for characterizing a wide range of 3D topological insulators, in which the contribution of conducting surface states has not been studied yet. 

 A schematic diagram of the coaxially-aligned double Corbino disk geometry is shown in Fig.~(\ref{Fig:DoubleSidedTwoTerminalCorbino}). This transport geometry can be realized by two steps of lithography performed on opposite sides of the sample. One surface (since both surfaces are identical) of the sample is shown in Fig.~(\ref{Fig:DoubleSidedTwoTerminalCorbino})~(a). The disk shapes are defined by highly conductive metal contacts (shown in yellow) used as terminals for resistance measurements. We note that the measurement configurations that will be discussed in the following will include both two- and four-terminal resistance measurements. 

 The two-terminal resistance measurements where the current and voltage leads share the same terminal should only be used when the contact resistances are negligible. One example of the two-terminal resistance measurement is shown in Fig.~(\ref{Fig:DoubleSidedTwoTerminalCorbino})~(b), where resistance is measured between the inner-metallic circle (terminal 1) and the outer-metallic region (terminal 2), or $R_L=V_{1,2}/I_{1,2}$. When the current flows only on the surface ($\sigma_{b}=0$ and $\sigma_{s}\neq0$), this is identical to Eq.~(\ref{Eq:SurfCorbinoFormula4}): $R_{L}=R_{\mathrm{Corbino}}$. 
 
 The two-terminal resistance can also be measured using both the top and bottom Corbino disks in parallel using the radial configuration as shown in Fig.~(\ref{Fig:DoubleSidedTwoTerminalCorbino})~(c). In the radial configuration, terminals 1 (2) and 3 (4) are connected with a low resistance jumper wire shown in blue (red). These jumper wire connections ensure that the top and bottom surfaces have identical electric potential profiles. In addition, the wire that connects terminals 2 and 4 (shown in red) prevents the current from flowing on the side surfaces of the sample. When $\sigma_{b}=0$ and $\sigma_{s}\neq0$, the radial resistance is equivalent to two resistors, corresponding to each Corbino lateral resistance measurement, connected in parallel: 
\begin{equation}
	R_{R}=\frac{1}{2}\times\frac{1}{2\pi}\ln(\frac{r_{\mathrm{out}}}{r_{\mathrm{in}}})\frac{1}{\sigma_{s}}.
	\label{Eq:RadialTwoTerminalDoubleCorbino}
\end{equation}

 Another two-terminal resistance measurement, which we will call the vertical configuration, can be performed on such a device as shown in Fig.~(\ref{Fig:DoubleSidedTwoTerminalCorbino})~(d). This configuration also shorts terminals 2 and 4 with a low resistance jumper wire (shown in red), which eliminates the current flow on the side surfaces. The vertical resistance is measured between the two inner-metallic circles, or $R_{V}=V_{1,3}/I_{1,3}$. In the bulk-dominated regime ($\sigma_{b}t\gg\sigma_{s}$), the current chooses a vertical path through the bulk between the two circular plates 1 and 3. In the surface-dominated regime ($\sigma_{b}t\ll\sigma_{s}$), the current flows first radially on the surface through the top Corbino disk, then passes through the jumper wire, and finally converges radially through the bottom Corbino disk. The approximate functional forms of resistance can be found in those two limits. In the $\sigma_{b}t\gg\sigma_{s}$ case, $R_{V}$ can be found by a derivation analogous to finding the capacitance between two parallel plates. If the sample is thin, we can assume the current density is in the vertical direction and uniform in the region of center contacts, so the resistance is:
\begin{equation}
	R_{V}\approx\frac{t}{\pi r_{1}^2}\frac{1}{\sigma_{b}},
	\label{Eq:VerticalTwoTerminalDoubleCorbino_BulkDom}
\end{equation} 
where $t$ is the thickness of the sample. In the other extreme, $\sigma_{b}t\ll\sigma_{s}$, we can regard each Corbino disks as resistors that are connected in series: 
\begin{equation}
	R_{V}=2\times\frac{1}{2\pi}\ln(\frac{r_{\mathrm{out}}}{r_{\mathrm{in}}})\frac{1}{\sigma_{s}}.
	\label{Eq:VerticalTwoTerminalDoubleCorbino_SurfDom}
\end{equation}  
Especially in the vertical measurement, because $R_{V}$ can be dramatically different in the two extremes, $\sigma_{b}t\gg\sigma_{s}$ and $\sigma_{b}t\ll\sigma_{s}$, and can be evaluated without numerical simulations, this configuration can provide strong evidence for experiments that are in the stage of verifying the existence of the surface states.  

 The last configuration that we consider is a four-terminal measurement, which we will call it the hybrid resistance measurement, as shown in Fig.~(\ref{Fig:DoubleSidedTwoTerminalCorbino})~(e).  In the hybrid configuration, the current flows between terminals 1 and 2, and the voltage is measured using terminals 3 and 4, which are on the opposite side, i.e., $R_{H}=V_{3,4}/I_{1,2}$. Since current lead 2 fully encloses lead 1, the hybrid resistance is an inverted resistance measurement. Because of the proximity of the voltage and current contacts, the inverted resistance signal in this double-sided device is expected to be much larger than that of the single-sided inverted resistance measurement discussed previously. 
\begin{figure}[t]
\begin{center}
\includegraphics{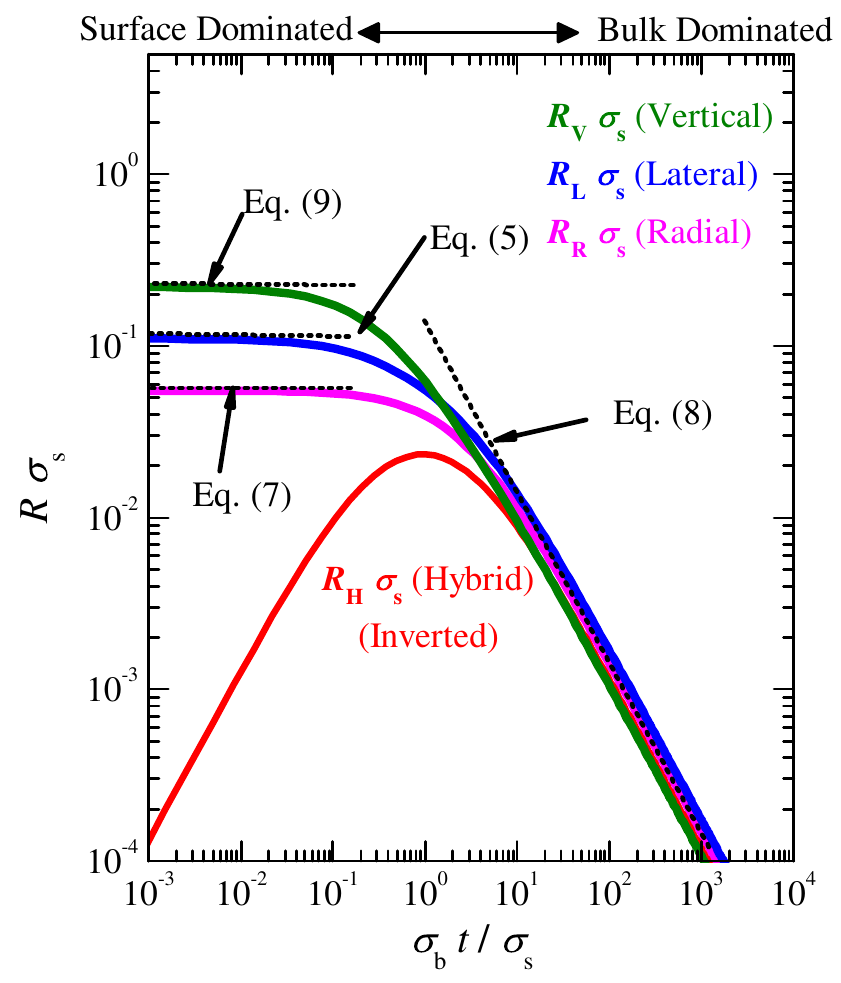}
\caption[Example of numerical results for a two-terminal double-sided Corbino disk.]{Example of numerical results for a two-terminal double-sided Corbino disk. Numerical result of $R_{L}\sigma_{s}$ (blue), $R_{R} \sigma_{s}$ (magenta), $R_{V}\sigma_{s}$ (green) and $R_{\mathrm{Inv}}\sigma_{s}$ (red) vs. $x$ ($=\sigma_{b}t/\sigma_{s}$) when the thickness is $t$ = 100 $\mathrm{\mu m}$. The dotted lines indicate the asymptotic values from the equations.}
\label{Fig:DoubleCorbinoTwoTermExampleResult}
\end{center}
\end{figure}
 We have solved this geometry numerically for different resistance configurations using finite element analysis (Comsol Multiphysics AC/DC module). In Fig.~(\ref{Fig:DoubleCorbinoTwoTermExampleResult}), we plot the dimensionless resistance $f$ ($=R\sigma_{s}$) for all four resistances as a function of $x$ ($=\sigma_{b}t/\sigma_{s}$). Indeed, as we expect, our hybrid resistance results in, $R_{H}\sigma_{s}\propto x$ when $x\rightarrow0$ in the inverted resistance measurement. In the opposite limit when $x\rightarrow\infty$ , $R\sigma_{s}\propto1/x$ for all four resistances as expected. In particular, $R_{V} \sigma_{s}$ approaches to Eq.~(\ref{Fig:DoubleCorbinoTwoTermExampleResult}). Furthermore, the asymptotic flat lines in $R_{L}\sigma_{s}$, $R_{R}\sigma_{s}$, and $R_{V}\sigma_{s}$ when $x\rightarrow0$ agrees with Eq.~(\ref{Eq:SurfCorbinoFormula4}) ($C_{0}=0.0645$), Eq.~(\ref{Eq:RadialTwoTerminalDoubleCorbino}) ($C_{0}=0.032$), and Eq.~(\ref{Eq:VerticalTwoTerminalDoubleCorbino_SurfDom}) ($C_{0}=0.129$), respectively. 
\begin{figure}[p]
\begin{center}
\includegraphics[scale=0.8]{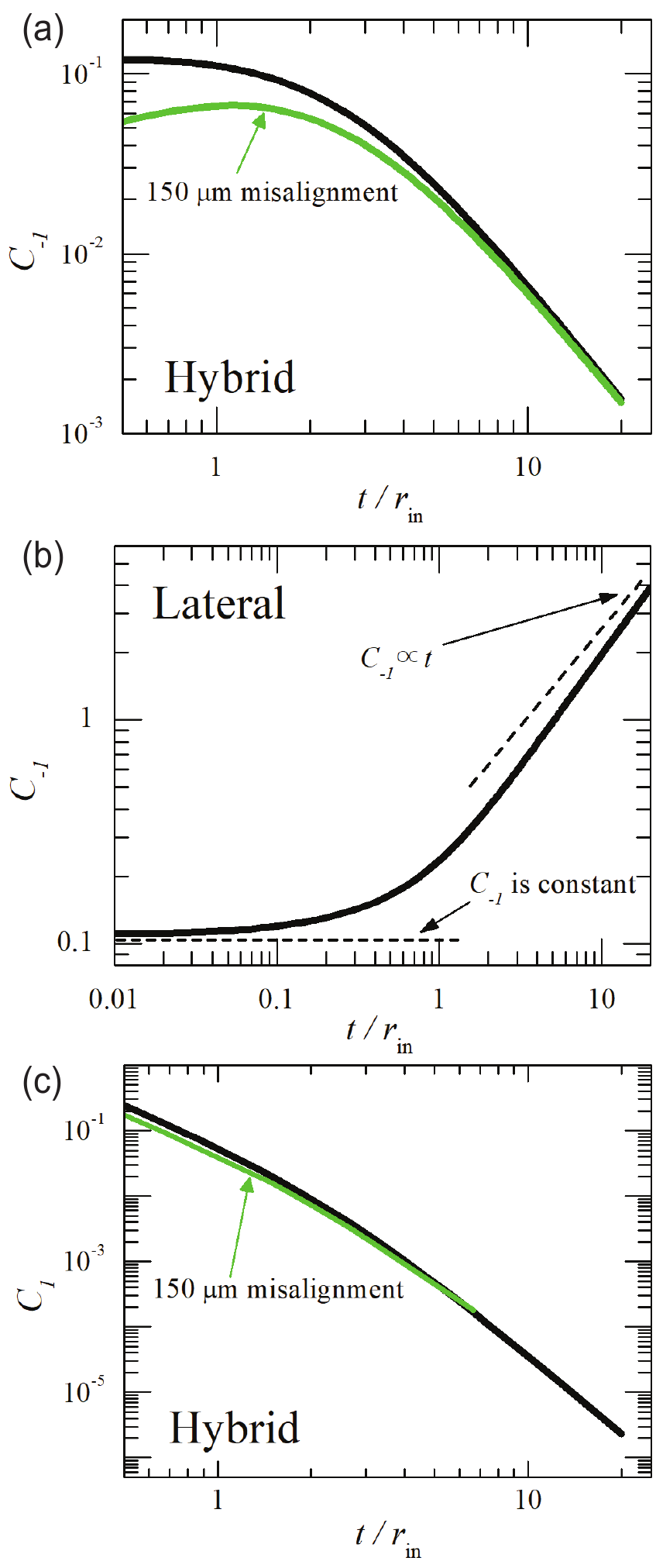}
\caption[The transport coefficients for the double-sided two-terminal Corbino disk geometry.]{The transport coefficients for the double-sided two-terminal Corbino disk geometry by varying the sample thickness. The solid black line shows the transport coefficient when the two Corbino disks are coaxially aligned, and the solid green line shows the transport coefficients when the two Corbino disks are misaligned by 150 $\mathrm{\mu m}$. The thickness is expressed in a dimensionless form by dividing by $r_{in}$ (= 150 $\mathrm{\mu m}$). (a) $C_{-1}$ of the hybrid resistance, $R_{H}$, in the bulk-dominated regime ($\sigma_{b}t\gg\sigma_{s}$). (b) $C_{-1}$ of the lateral resistance, $R_{L}$, in the bulk-dominated regime ($\sigma_{b}t\gg\sigma_{s}$). (c) $C_{1}$ of the hybrid resistance, $R_{H}$, in the surface-dominated regime ($\sigma_{s}\gg\sigma_{b}t$).}
\label{Fig:DoubleCorbinoTwoTermCoefficient}
\end{center}
\end{figure}
 We have also solved $C_{-1}$ and $C_{1}$ iteratively for different thicknesses. In Fig.~(\ref{Fig:DoubleCorbinoTwoTermCoefficient}), we present our results for $C_{-1}$ and $C_{1}$ as a function of dimensionless thickness ($t/r_{1}$). In experiments, $R_{L}$, $R_{R}$, and $R_{V}$ may suffer from the presence of contact resistances. To address this potential problem, we present $C_{-1}$  from the $R_{H}$, which is a four-terminal measurement. Comparing to Fig.~(\ref{Fig:DoubleCorbinoTwoTermCoefficient})~(a) and Fig~(\ref{Fig:TransCoeffSingleCorbino})~(a), when the sample is thin ($t<r_{1}$), $C_{1}$ from the hybrid measurement is a few orders of magnitude larger than the inverted measurement from the single-sided 4-terminal Corbino disk case. Therefore, this transport geometry can be better in extracting a smaller bulk conductivity.
 
 However, there can be a practical difficulty when preparing a double-sided transport geometry: the two Corbino disks may be misaligned. This type of misalignment would be expected to change the values of $C_{-1}$ and $C_{1}$. To understand how the misalignment impacts the resistance measurements, we have repeated the calculations for the case where there is a 150 $\mathrm{\mu m}$ misalignment between the top and bottom Corbino disks. The coefficients $C_{-1}$ and $C_{1}$ calculated with such a misalignment are shown in green. Our numerical calculations indicate that the hybrid resistance measurement is most vulnerable to misalignment. In the following subsection, we consider a transport geometry that allows similar measurements, but all four measurements are in a 4-terminal configuration.
 
\subsection{Double-Sided Four-terminal Double Corbino Disk}

\begin{figure}[t]
\begin{center}
\includegraphics[scale=0.9]{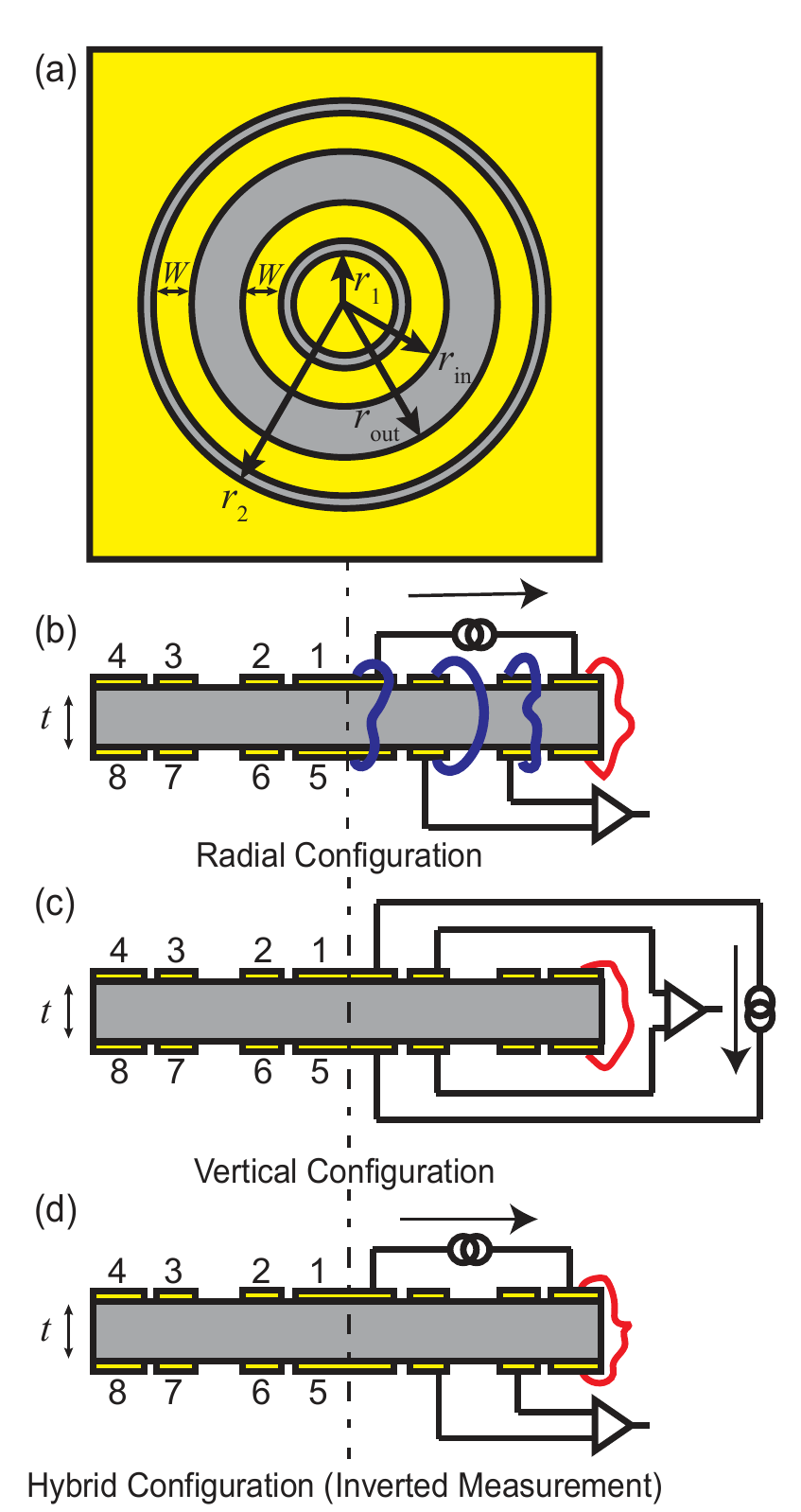}
\caption[Four-terminal double-sided Corbino disk.]{Four-terminal double-sided Corbino disk. The sample is shown in gray and the highly conductive contacts are shown in yellow. (a) Top and bottom view of the sample. (b) Side view of the sample and the radial configuration. (c) Side view of the sample and the vertical configuration. We choose the Corbino disk dimensions identical to the single-sided four-terminal Corbino disk shown in Fig.~(\ref{Fig:FourTermCorbino}): $r_{1}$ = 100 $\mathrm{\mu m}$, $r_{2}$ = 800 $\mathrm{\mu m}$, $r_{in}$ = 200 $\mathrm{\mu m}$, $r_{out}$ = 300 $\mathrm{\mu m}$, $W$ = 75 $\mathrm{\mu m}$.}
\label{Fig:DoubleCorbinoFourTerm}
\end{center}
\end{figure}

 The single-sided four-terminal Corbino disk geometry (in subsec.~\ref{subsec:SingleFourTerminalInv}) has the limitation in that the $R_{\mathrm{Inv}}$ may be too small to measure in the surface-dominated regime ($\sigma_{b}t\ll\sigma_{s}$). The coaxially aligned double Corbino disk geometry (in subsec.~\ref{subsection:DoubleSidedInvert}) can give a larger magnitude from the $R_{H}$ if the sample is thin, but the other three resistances ($R_{L}$, $R_{R}$, and $R_{V}$) may suffer from contact resistances in the bulk-dominated regime ($\sigma_{b}t\gg\sigma_{s}$). In this subsection, we consider a transport geometry that incorporates advantages of the two geometries, so that both issues can be eliminated. The configuration consists of two coaxially aligned four-terminal Corbino disks placed on both sides of the sample as illustrated in Fig.~(\ref{Fig:DoubleCorbinoFourTerm}). The advantages of having multiple leads on both sides should be clear: we now have more than enough 4-terminal resistance configurations, including inverted measurements, that can be used to extract the bulk conductivity. Even without using any jumper wires, such an 8-terminal transport geometry would allow us to perform a total of 70 independent inverted resistance measurements. We will only discuss a few that are similar to what was discussed in the previous sections. In addition, we will provide an example of how to apply a ratio method to extract the two conductivities. 
  A schematic diagram of the sample is shown in Fig.~(\ref{Fig:DoubleCorbinoFourTerm})~(a). The disk shapes are defined by highly conductive metal contacts (shown in yellow). This type of sample can be realized by two separate steps of lithography. The conductive metal regions are used as terminals for resistance measurements. We choose the dimensions of the Corbino disks with two rings to be identical to the earlier Corbino disk structure shown in Fig.~(\ref{Fig:FourTermCorbino}). Therefore, the single-sided 4-terminal Corbino disk results can be used for this geometry as well. Because of this, we will omit the resistance configurations that uses contacts from a single side, and instead we will discuss three new measurement configurations that require the contacts from both top and bottom surfaces. 

 We first discuss the radial measurement configuration ($R_{R}$) shown in Fig.~(\ref{Fig:DoubleCorbinoFourTerm})~(b). Terminal pairs of (1, 5), (2, 6), and (3, 7) are connected with a low-resistance jumper wire shown in blue. These jumper wire connections help to enhance the radial flow of current in the bulk. In addition, the wire that connects terminals 4 and 8 (shown in red) prevents the current from flowing on the side surfaces of the sample. When the current only flows on the surface ($\sigma_{s}\neq0$ and $\sigma_{b}=0$), the radial resistance is:  
\begin{equation}
	R_{R}=\frac{1}{2}R_{L}=\frac{1}{2}\times\frac{1}{2\pi}\ln(\frac{r_{\mathrm{out}}}{r_{\mathrm{in}}})\frac{1}{\sigma_{b}},
	\label{Eq:RadialFourTerminalDoubleCorbino_Surface}
\end{equation} 
where $R_{L}$ is the lateral resistance that was introduced in Eq.~(\ref{Eq:SurfCorbinoFormula4}).

 Next, we consider the vertical configuration shown in Fig.~(\ref{Fig:DoubleCorbinoFourTerm})~(c). Similar to the radial configuration, terminals 4 and 8 are connected with a low resistance jumper wire shown in red in order to eliminate the current flowing on the side surfaces. In the bulk-dominated regime ($\sigma_{b}t\gg\sigma_{s}$), current flows mostly vertically through the bulk. The resistance is thickness dependent and can be expressed in the form of $R_{V}\sigma_{s}\approx C_{-1}(\sigma_{b} t/\sigma_{s} )^{-1}$, which is the leading order term of Eq.~(\ref{Eq:BulkDomSeries}). In the surface-dominated regime ($\sigma_{b}t\ll\sigma_{s}$), where the current flows mostly on the surfaces, first the current flows radially outwards on the top surface, and then the current flows to the other surface through the jumper wire and converges radially inwards. In this case, we can regard every possible conduction path of the Corbino disks as a resistor and add them in series: 
 \begin{equation}
	R_{V}=2\times\frac{1}{2\pi}\ln(\frac{r_{\mathrm{out}}}{r_{\mathrm{in}}}\frac{r_{2}}{r_{1}}\frac{r_{\mathrm{in}}-W}{r_{\mathrm{out}}+W})\frac{1}{\sigma_{s}}.
	\label{Eq:VerticallFourTerminalDoubleCorbino_Surface}
\end{equation} 

 For the hybrid configuration, as shown in Fig.~(\ref{Fig:DoubleCorbinoFourTerm})~(d), while terminals 4 and 8 are connected with a low resistance jumper wire (shown in red) to ensure that the side surface contribution is eliminated, the current leads are connected to terminals 1 and 4, and the voltage leads are connected to terminals 6 and 7, i.e., $R_{H}=V_{6,7}/I_{1,4}$. When the current chooses its path mostly on the surface ($\sigma_{b}t\ll\sigma_{s}$), this is indeed the inverted measurement in that the Corbino disk on the top surface corresponds to the enclosed loop that captures the current flowing on the surface and the bottom Corbino disk corresponds to the voltage measurement external to that loop and measures the current path contributing from the bulk.
\begin{figure}[t]
\begin{center}
$\hspace*{-1cm}$
\includegraphics[scale=0.9]{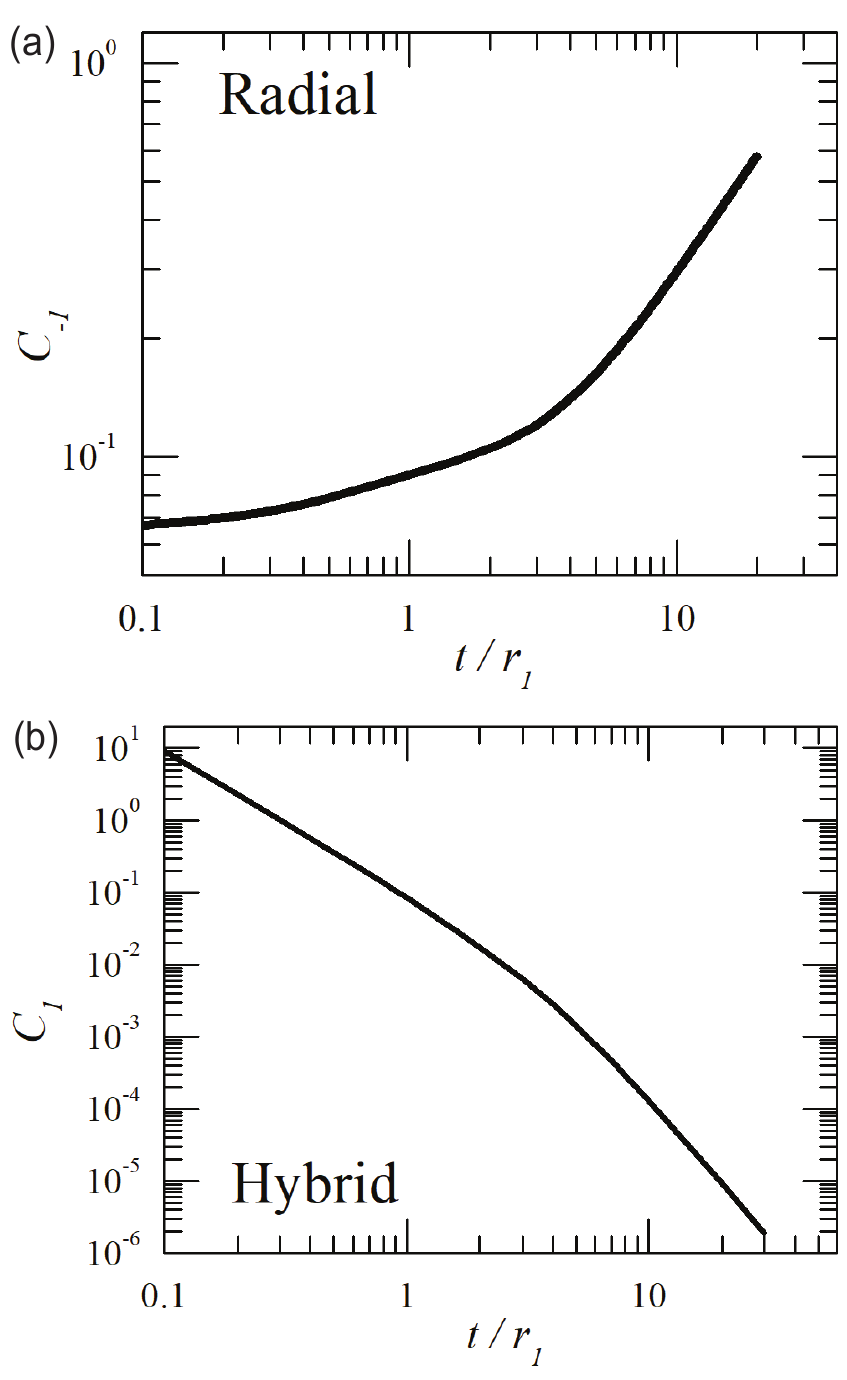}
\caption[Transport coefficients for the four-terminal double Corbino disk geometry.]{Transport coefficients for the four-terminal double Corbino disk geometry at different thicknesses. (a) $C_{-1}$ for the radial resistance, $R_{R}$, in the bulk-dominated regime ($\sigma_{b}t\gg\sigma_{s}$). (b) $C_{-1}$ for the hybrid resistance, $R_{H}$, in the surface-dominated regime ($\sigma_{b}t\gg\sigma_{s}$). The thickness is expressed in a dimensionless form by dividing by $r_{1}$ (= 100 $\mathrm{\mu m}$).}
\label{Fig:CoefficientFourTerminalDoubleCorbino}
\end{center}
\end{figure}
  Similar to the previous transport geometries discussed above, we calculated the relevant transport coefficients for this sample geometry numerically using finite element analysis (Comsol Multiphysics AC/DC module). We present $C_{1}$ for the hybrid measurements and $C_{-1}$  for radial measurements as a function of thickness in Fig.~(\ref{Fig:CoefficientFourTerminalDoubleCorbino}). If the top and bottom Corbino rings are significantly coaxially misaligned, we recommend using $C_{-1}$ from the lateral measurements (Fig.~(\ref{Fig:TransCoeffSingleCorbino})~(a)) instead of using $C_{-1}$ from the radial measurement. 
 
\section{\label{sec:Realization}Experimental Realization of the Transport Geometries}

 In this section, we demonstrate the transport geometries introduced in Sec.~\ref{sec:TransportGeometry} experimentally on samarium hexaboride (SmB$_{6}$). SmB$_{6}$ is a cubic material, and it appears to be a topological Kondo insulator with a truly insulating bulk and robust surface conduction on all surfaces. For typical samples, the transport is dominated by bulk conduction above $\sim$4 K. Previous transport experiments indicate that the bulk resistivity of SmB$_{6}$ is activated in the form of $R$ $\propto \exp$( $E_{a}$/$k_{B}T$), with an activation energy, $E_{a}$, of about 3 $\textendash$~4 meV. The activated behavior of the bulk transport is difficult to study using standard transport experiments below 4 K, because at that temperature, current starts to flow mostly through the surface. In contrast, the inverted resistance measurements can be used to measure the bulk conductivity of SmB$_{6}$ at temperatures below 4 K. We will use the single-sided 4-terminal Corbino disk that was introduced in subsection~\ref{subsec:SingleFourTerminalInv}, and the double-sided 2-terminal Corbino disk that was introduced in subsection~\ref{subsection:DoubleSidedInvert} to characterize bulk and surface transport in SmB$_{6}$. Consistent with our expectations, we were able to measure bulk conductivity below the bulk-to-surface crossover temperature. A more detailed study of the underlying physics of bulk transport in hexaborides in this new range of temperatures will be discussed in a later study\cite{YSEo}.
 
\subsection{\label{subsec:singlesideresult}Single-Sided Four-Terminal Corbino Disk on SmB$_{6}$} 
 We prepared a transport geometry on a polished SmB$_{6}$ surface using the dimensions shown in Fig.~(\ref{Fig:FourTermCorbino}). The SmB$_{6}$ crystal used in this study was grown in an aluminum flux. The grown crystal was first thinned with a SiC polishing pad that has grit size of 15.3 $\mathrm{\mu m}$. It was then polished with a SiC polishing pad of 2.5 $\mathrm{\mu m}$. Fine polishing was performed on polishing cloth (TexMet C) using an aluminum oxide slurry with a particle size of 0.3 $\mathrm{\mu m}$. After cleaning the surface with diluted HCl, we performed photolithography on the polished surface with a mask design of the 4-terminal Corbino disk pattern. We exposed the surface where we wanted to evaporate a highly conductive metal, and covered it with photoresist on the regions that we want to expose later. We then evaporated Ti/Au 20 $\mathrm{\AA}$/1500 $\mathrm{\AA}$, and lifted off the photoresist in acetone. The sample was cleaned again with diluted HCl, and then attached to a silicon piece with an insulating SiO$_{2}$ layer using Torrseal. We used a wire bonder to attach aluminum wires (1 mil) to the terminals. We performed low-frequency resistance measurements at temperatures ranging from 16 to 1.7 K using the quantum design PPMS system connected to an external lock-in amplifier (SR830). 
\begin{figure}[p]
\begin{center}
\includegraphics[scale=0.9]{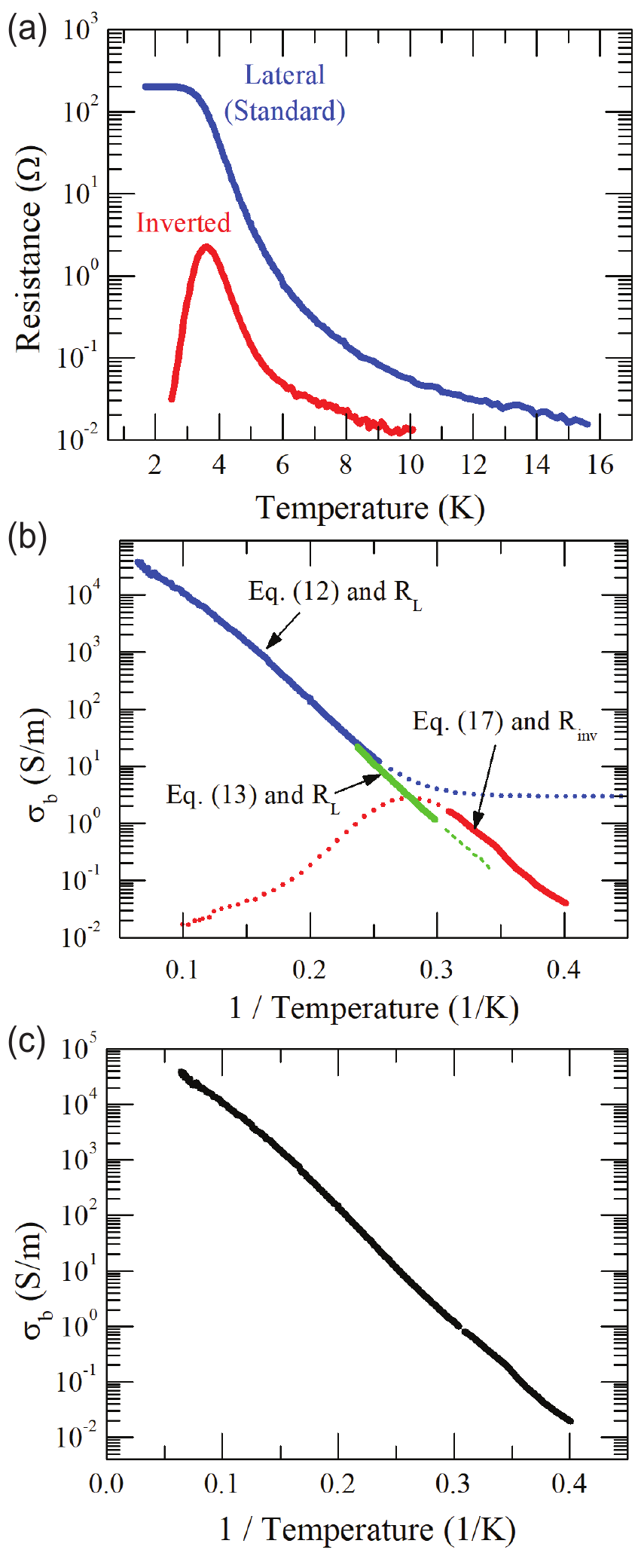}
\caption[Experimental results and bulk conductivity analysis on a single-sided 4-terminal Corbino disk on a SmB$_{6}$ sample.]{Experimental results and bulk conductivity analysis on a single-sided 4-terminal Corbino disk on a SmB$_{6}$ sample with thickness, $t$ = 300 $\mathrm{\mu m}$. (a) Resistance vs. temperature of the lateral and inverted measurements. (b) The bulk conductivity converted from the measured resistance. The dotted lines indicate the bulk conductivity conversion applied beyond the appropriate regime. (c) The result of the extracted bulk conductivity after adjustment of mismatch.}
\label{Fig:SingleCorbinoExpResult}
\end{center}
\end{figure}

The experimental results, plotted as resistance vs$.$ temperature, are shown in Fig.~(\ref{Fig:SingleCorbinoExpResult})~(a). As temperature is decreased from 16 K, both the lateral resistance, $R_{L}$ (shown in blue), and the inverted resistance, $R_{\mathrm{Inv}}$ (shown in red), increases. In this temperature range, the sample is in the bulk-dominated regime ($\sigma_{b}t\gg\sigma_{s}$). Then around 3.5 K, $R_{\mathrm{Inv}}$ reaches a peak and starts to decrease, while $R_{L}$ plateaus. The sample at this low temperature is in the surface-dominated regime ($\sigma_{b}t\ll\sigma_{s}$). According to Eq.~(\ref{Eq:SurfCorbinoFormula4}), the surface conductivity is $\sigma_{s}=3.3\times10^{-4}$ S. Meanwhile, $R_{\mathrm{Inv}}$ continuously drops toward 0 Ω until it is too noisy to measure. The resolution of our electronics allowed us to measure $R_{\mathrm{Inv}}$ down to $\sim$2.5 K. 

To extract the bulk conductivity from the data in Fig.~(\ref{Fig:SingleCorbinoExpResult})~(a), we use leading order terms in the series expansion in the surface- and bulk-dominated regimes (Eq.~(\ref{Eq:BulkDomSeries}) \textendash ~Eq.~(\ref{Eq:InvSeries})), and the two-channel model (Eq.~(\ref{Eq:CorbinowithBulk})). For each temperature range, we converted the resistance measurements to bulk conductivity as described in the following.

\textit{High temperature range} (bulk-dominated regime): At temperatures above $\sim$3.6 K, the lateral resistance can be used to find the bulk conductivity. In this bulk-dominated regime ($\sigma_{b}l\gg\sigma_{s}$), Eq.~(\ref{Eq:BulkDomSeries}) can be used to understand the resistance behavior (we replaced $l$ with thickness, $t$). The contribution of the current that flows on the surface is extremely small, so we keep only the first order term. With this first order term, and using $C_{-1}$ found in Fig.~(\ref{Fig:TransCoeffSingleCorbino})~(a), the bulk conductivity is:
\begin{equation}
	\sigma_{b}(T)=\frac{C_{-1}}{t}\frac{1}{R_{L}(T)}.
	\label{Eq:HighTempBulkCond_SingleCorbino}
\end{equation}
The result is shown in Fig.~(\ref{Fig:SingleCorbinoExpResult})~(b) as a solid blue line. 

\textit{Intermediate temperature range} (bulk-to-surface crossover regime): In the temperature range, 3.3 - 3.6 K, the sample undergoes a bulk-to-surface crossover, so the bulk and surface conductions are comparable. This regime is where the expansion of the $f(x)$, cannot be expanded in series in either extremes of $x$ because $x\approx1$. Instead, we make use of Eq.~(\ref{Eq:CorbinowithBulk}) from the two-channel model to convert $R_{L}$ to bulk conductivity. This can be re-formulated as:
\begin{equation}
	\sigma_{b}(T)=\frac{1}{\gamma}(\frac{C_{0}}{R_{L}(T)}-\sigma_{s}).
	\label{Eq:IntermTempBulkCond_SingleCorbino}
\end{equation}
The result is shown in Fig.~(\ref{Fig:SingleCorbinoExpResult})~(b) as a solid green line. We have shown in  Fig.~(\ref{Fig:StandTwoTermCorbinoResult}) that the calculated effective thickness is slightly different for the bulk-dominated and the surface-dominated regimes.  In this particular experiment, we estimate the effective thickness from $\gamma=C_{0}t/C_{-1}$. It is fortunate that the temperature dependence of the surface conductivity in SmB$_{6}$ is weak enough that it can be approximated as constant. 

We warn the reader that Eq.~(\ref{Eq:IntermTempBulkCond_SingleCorbino}) should not be used alone in samples with a surface conductivity that is strongly temperature dependent. For characterizing such a material, in this intermediate temperature range, it is important to use multiple four-terminal resistance measurements that put different emphasis on surface and bulk conductivity, Luckily the double-sided Corbino geometry allows one to perform a vertical measurement ($R_{V}$), which would put greater emphasis on bulk conductivity. Thus, the combination of lateral and vertical measurements may provide a better strategy for extracting the bulk conductivity in this temperature regime.

We expect that the process of extracting the conductivities from the measured resistances would have been much more challenging if the surface conductivity was also strongly temperature dependent. In such a case, the full numerical simulation must be done, and then compared to the measured resistances. 

According to Eq.~(\ref{Eq:ScaleResistance}), one can measure the resistance, $R$, easily, but not $f(x)$ if $\sigma_{s}$ has a strong temperature dependence. One may implement an approach that would rely on the ratios of the two different resistance measurements. For example, if we take the ratio of the conventional resistance, $R_{C}$, to the inverted resistance, $R_{\mathrm{Inv}}$, we can directly compare to the ratios of the dimensionless functions, $f(x)$, of the corresponding resistances, $f_{C}(x)$ and $f_{\mathrm{Inv}}(x)$:
\begin{equation}
	\frac{R_{C}}{R_{\mathrm{Inv}}}=\frac{f_{C}}{f_{\mathrm{Inv}}}=h(x).
	\label{Eq:ScaleRatio_SingleCorbino}
\end{equation}
where the $f_{C}(x)$ is the dimensionless resistance for $R_{C}$, $f_{\mathrm{Inv}}(x)$ is the dimensionless resistance for $R_{\mathrm{Inv}}$, and we define the ratio as $h(x)$. From Eq.~(\ref{Eq:ScaleRatio_SingleCorbino}), we can compare the experimental data and numerical results directly, and find $x$. The surface conductivity can then be found by:
\begin{equation}
	\frac{R_{C}}{f_{C}}=\frac{1}{\sigma_{s}} \quad \mathrm{or} \quad \frac{R_{\mathrm{Inv}}}{f_{\mathrm{Inv}}}=\frac{1}{\sigma_{s}}.
	\label{Eq:ScaleRatio_SingleCorbino2}
\end{equation}
After $\sigma_{s}$ is found, $\sigma_{b}$ can be found from $x$ that was found from Eq.~(\ref{Eq:ScaleRatio_SingleCorbino}):
\begin{equation}
	\sigma_{b}=\frac{\sigma_{s}x}{t}.
	\label{Eq:ScaleRatio_SingleCorbino3}
\end{equation}

\textit{Low temperature range} (surface-dominated regime): For temperatures below $\sim$3.3 K, the sample is in the surface-dominated regime. The inverted resistance measurement can be used at this temperature range to extract the bulk conductivity. In the surface-dominated regime, the inverted resistance can be expressed by Eq.~(\ref{Eq:InvSeries}). $C_{1}$ can be found from Fig.~(\ref{Fig:TransCoeffSingleCorbino})~(b), and $\sigma_{s}$ can be found from Eq.~(~\ref{Eq:SurfCorbinoFormula4}) and the plateau value of $R_{L}(T)$. Then the bulk conductivity is:
\begin{equation}
	\sigma_{b}(T)=\frac{\sigma_{s}(T)^2}{t}\frac{R_{\mathrm{Inv}}}{C_{1}}.
	\label{Eq:LowTempBulkCond_SingleCorbino}
\end{equation}
The result is shown in Fig.~(\ref{Fig:SingleCorbinoExpResult})~(b) as a solid red line. 

Notice that there is a noticeable mismatch of about a factor of $\sim$3 between the two plots (solid blue line and the solid red line) near the bulk-to-surface crossover in Fig.~(\ref{Fig:SingleCorbinoExpResult})~(b). We have also tested five other SmB$_{6}$ samples, and found mismatch factors ranging from 0.85\textendash ~3.5. The mismatch indicates that the geometric coefficients ($C_{-1}$, $C_{1}$, etc.) do not correspond to the actual sample geometry. There are numerous possibilities that may have caused this mismatch. For SmB$_{6}$, one should worry about the quality difference between the top and bottom surface. This can greatly influence the inverted resistance measurement since the $R_{\mathrm{Inv}}\propto 1/\sigma_{s}^2$. Aluminum inclusions that are known to be possibly present in flux grown samples can influence the current path in the bulk, and result in a different geometric coefficient. 

Even in the case where the top and bottom surfaces have identical transport properties and the sample is free from inclusions, a slight difference of dimensions between the realized sample geometry and the numerically-simulated geometry can result in a large mismatch as above, especially for the inverted resistance measurement. The inverted resistance measurement depends greatly on the fringe currents in the bulk, and measurements that involve these current paths are highly sensitive to the details of the geometry. For example, a similar discrepancy also occurs in conventional transport geometries such as van der Pauw measurements. In van der Pauw measurements performed on cleaved square samples, they typically result in different 4-terminal resistance values depending on the direction of the current paths as in vertical and horizontal resistance measurements. The significant difference of these two measurements arises from small deviations from a perfect square shape of the sample. In fact, to obtain the true resistivity of the material, one needs to take the average value of the two measurements. 

The discrepancy of the bulk conductivity obtained from the inverted and lateral resistance measurements is similar to the discrepancy that occurs from the horizontal and vertical resistance measurements obtained from a van der Pauw measurement. Since the inverted resistance measurement is much more vulnerable to imperfections of the sample geometry, we recommend adjusting the $C_{1}$ value in such a way that inverted measurement matches the lateral measurement at around 3.3 K. 

After adjusting and combining the three plots in Fig.~(\ref{Fig:SingleCorbinoExpResult})~(b), we obtain the bulk conductivity that covers the entire temperature range shown in Fig.~(\ref{Fig:SingleCorbinoExpResult})~(c). From this exercise, we have demonstrated that the bulk conductivity can be found even in the surface-dominated regime. Also, we have shown that the bulk conductivity of SmB$_{6}$ continues to exhibit a thermally excited behavior according to $\sigma_{b}\propto \exp(-E_{a}/k_{B}T)$, where we find the activation behavior of $E_{a} =3.84$ meV. 

\subsection{Double-Sided Two-Terminal Corbino Disk on SmB$_{6}$}

In the previous subsection, we measured the inverted resistance, $R_{\mathrm{Inv}}$, successfully down to 2.5 K. According to Eq.~(\ref{Eq:CorbinowithBulk}), the magnitude of the $R_{\mathrm{Inv}}$ is inversely proportional to $\sigma_{s}^2$. For samples that have a higher surface conductivity, the magnitude of $R_{\mathrm{Inv}}$ will be smaller in general, and therefore the measurement may fail at higher temperatures. To overcome these limitations, we instead use a transport geometry that allows larger $C_{1}$ values. 

In this subsection, we demonstrate the bulk conductivity extraction from the coaxially aligned two-terminal double Corbino disk geometry that is shown in Fig.~(\ref{Fig:DoubleSidedTwoTerminalCorbino}). We use SmB$_{6}$ crystals from the same batch to make this double-sided sample. The top and bottom surfaces of the sample were first polished by using a SiC polishing pad of 15.3 $\mathrm{\mu}$m. Then, the surfaces were polished by finer grit sizes of SiC pads: 5.5 $\mathrm{\mu}$m and 2.5 $\mathrm{\mu}$m. The polishing was finalized by using a polishing cloth (TexMet C) with 0.3 $\mathrm{\mu}$m Al$_{2}$O$_{3}$ slurry. We performed photolithography on each of the polished surfaces with a mask design with identical dimensions to the two-terminal Corbino disks shown in Fig.~(\ref{Fig:DoubleSidedTwoTerminalCorbino})~(a). We evaporate Ti/Au (20 $\mathrm{\AA}$ /1500 $\mathrm{\AA}$), and then lifted off the photoresist on the regions where we intend to expose the surface. We attached the sample with GE varnish on a silicon piece with native oxide so that both surfaces were exposed. We used copper wires (2 mils), and attached them with silver paste. 
\begin{figure}[p]
\begin{center}
\includegraphics[scale=0.9]{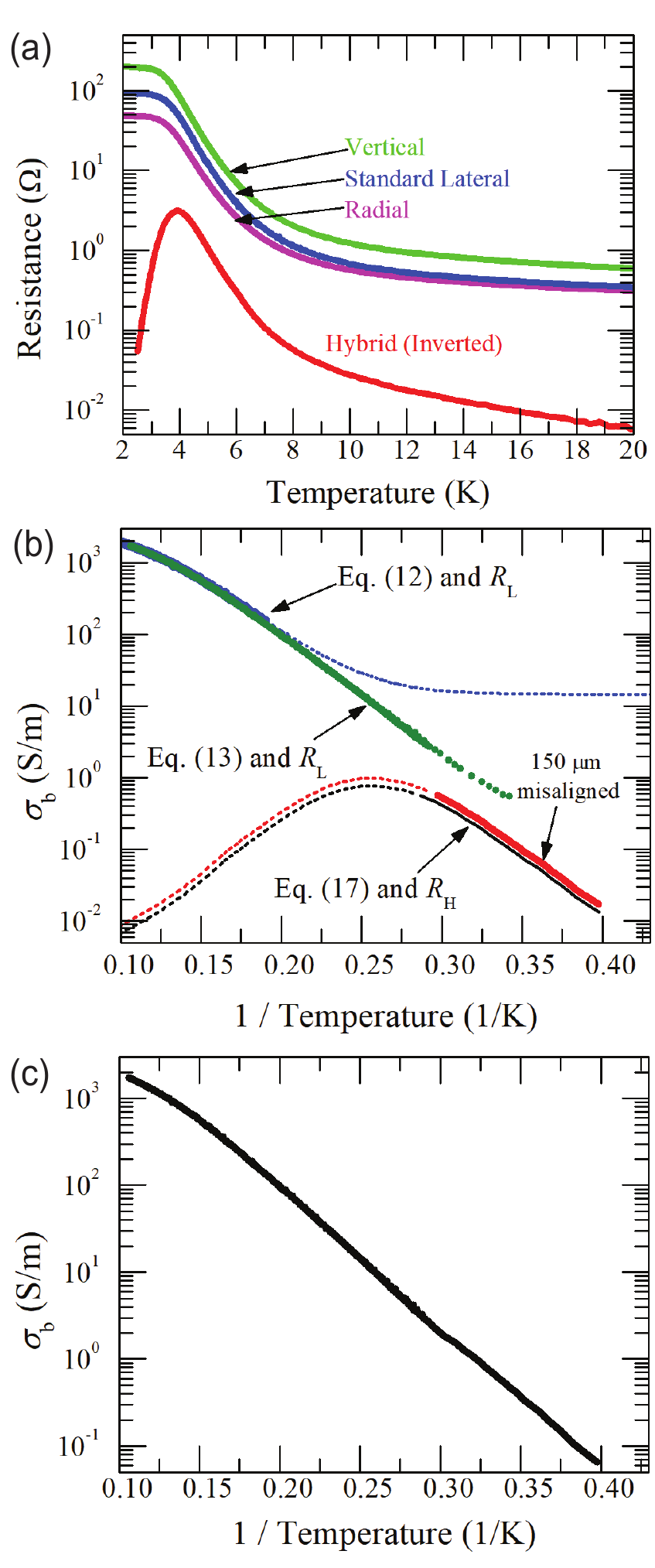}
\caption[Experimental results and bulk conductivity analysis of the double-sided two-terminal Corbino disk on a SmB$_{6}$ sample]{Experimental results and bulk conductivity analysis of the double-sided 2-terminal Corbino disk on a SmB$_{6}$ sample with thickness, $t$ = 210 $\mathrm{\mu m}$. (a) Resistance vs. temperature of the lateral, vertical, radial, and the hybrid measurement (inverted measurement). (b) The bulk conductivity converted from the measured resistance shown in (a). The dotted lines indicate the bulk conductivity conversion applied beyond the appropriate regime. (c) The result of the bulk conductivity after adjustment of mismatch.}
\label{Fig:DoubleCorbinoExpResults}
\end{center}
\end{figure}

The sample was characterized using standard low frequency lock-in measurements from room temperature to cryogenic temperatures (300 K \textendash ~2.5 K). We measured the resistances of the four different configurations shown in Fig.~(\ref{Fig:DoubleSidedTwoTerminalCorbino})~(b) \textendash ~(e) ($R_{L}$, $R_{R}$, $R_{V}$, and $R_{H}$). The experimental results of resistance vs temperature for all four measurements are shown in Fig.~(\ref{Fig:DoubleCorbinoExpResults})~(a). At temperatures below 3.6 K, in the surface-dominated regime, resistance plateaus are observed in $R_{L}$, $R_{R}$, and $R_{V}$. By comparing the magnitudes, the relations $R_{R}= 1/2 R_{L}$ (Eq.~(\ref{Eq:RadialTwoTerminalDoubleCorbino})) and $R_{V}\approx2R_{L}$ (Eq.~(\ref{Eq:VerticalTwoTerminalDoubleCorbino_SurfDom}) holds, which verifies the existence of the conducting surface. From, $R_{R}$, we find the surface conductivity, $\sigma_{s}=1.1\times10^{-3}$ S. This value is more than three times higher than the SmB$_{6}$ sample in the previous demonstration ($\sigma_{s}=3.3\times10^{-4}$ S). In the hybrid measurement, $R_{H}$, drops toward 0 $\mathrm{\Omega}$ as the temperature is lowered below 3.6 K. 

To extract the bulk conductivity from Fig~(\ref{Fig:DoubleCorbinoExpResults})~(a), we converted the resistance measurements to bulk conductivity by the following, similar to the single-sided 4-terminal Corbino disk case in the previous section.

\textit{High temperature range} (bulk-dominated regime): At high temperatures, above 5.2 K, the lateral resistance data is used to extract the bulk conductivity. Using $C_{-1}$ found in Fig.~(\ref{Fig:DoubleCorbinoTwoTermCoefficient})~(b) and Eq.~(\ref{Eq:HighTempBulkCond_SingleCorbino}), the bulk conductivity is found, as shown in the solid blue line in Fig.~(\ref{Fig:DoubleCorbinoExpResults})~(b).

\textit{Intermediate temperature range} (bulk-to-surface crossover regime): At intermediate temperatures, ranging from 3.5 K \textendash ~5.2 K, we again use the lateral resistance measurement and use Eq.~(\ref{Eq:IntermTempBulkCond_SingleCorbino}) with $\gamma=C_{0}t/C_{-1}$, which was derived from the two-channel model. The result is shown in the solid green line in Fig.~(\ref{Fig:DoubleCorbinoExpResults})~(b). 

\textit{Low temperature range} (surface-dominated regime):  At low temperatures, below $\sim$3.5 K, the sample is in the surface-dominated regime. Here, we use the hybrid resistance measurement to find the bulk conductivity. Eq.~(\ref{Eq:LowTempBulkCond_SingleCorbino}) can be used to convert the hybrid resistance to bulk conductivity. $\sigma_{s}$ can be found from the pleateau value of $R_{L}$, and $C_{1}$ can be found from Fig.~(\ref{Fig:DoubleCorbinoTwoTermCoefficient}). We used both the $C_{1}$ values when the top and bottom Corbino disks are perfectly aligned and when the two disks are misaligned by 150 $\mathrm{\mu m}$. The black solid line is the bulk conductivity when the two disks are perfectly aligned, the red solid line is the bulk conductivity when the two are misaligned by 150 $\mathrm{\mu m}$. 

Similar to the previous example, there is a mismatch between the bulk conductivity at low temperatures and the bulk conductivity at intermediate temperatures. The reason for this mismatch is identical to the reason for the mismatch present in the single-sided four-terminal demonstration in that the geometric coefficients may not correspond to the actual sample geometry. In addition to the reasons mentioned in the previous subsection, the misalignment of the top and bottom Corbino disks must also be considered. After the fabrication process, we find an unintentional misalignment of $\sim$150 $\mathrm{\mu m}$; however, even if we consider this misalignment, the corrected bulk conductivity still does not account the matching of the bulk conductivity found from the lateral resistance measurement at intermediate temperatures.  

Again, we adjust the bulk conductivity curve at low temperatures (solid red line) to match the conductivity at intermediate temperatures (solid green line). The result of bulk conductivity is shown in Fig.~(\ref{Fig:DoubleCorbinoExpResults})~(c). We note that in this particular sample the inverted resistance measurements allowed us to measure bulk conductivity down to 2.5 K when 99.9$\%$ of the current was flowing on the surface of the sample. We plan to present the bulk transport properties of SmB$_6$ and related materials and the implications of these measurements within the context of topological Kondo insulators in a separate publication\cite{YSEo}.

\section{\label{sec:Conclusion}Conclusion}

 We have introduced a new type of transport measurement, which we call the inverted resistance measurement. Together with a conventional transport measurement, it allows us to characterize materials that have both bulk and surface conduction, such as TIs. The inverted resistance measurement is powerful in the regime where the surface conduction dominates the bulk. The inverted resistance is proportional to $\sigma_{b}/\sigma_{s}^2$, and therefore the bulk conductivity, $\sigma_{b}$, is accessible even in the regime where surface conduction is dominant, making this measurement powerful. This inverted resistance measurement requires a loop as a current lead that encloses the other current lead, and two voltage leads placed outside of the current loop. We have analyzed different transport geometries that utilize this inverted measurement. The most ideal transport geometry for this inverted measurement is when two Corbino disks are coaxially aligned on two opposite surfaces. If the sample is thick, the inverted measurement can also be performed using a single-sided four-terminal Corbino disk geometry. We note that this type of measurement is not suitable for characterizing purely 2D systems or thin films as we expect the inverted resistance would be zero, regardless of the conductivities of the material. 

The geometric prefactors ($C_{-1}$ and $C_{1}$), which are used for converting from resistances to conductivities, were found using finite element analysis simulations. Experimentally, we have successfully realized the transport geometries on SmB$_{6}$ samples, and measured the resistances at different temperatures. SmB$_{6}$ turns out to be an ideal material for testing our transport method because the bulk has a thermally activated behavior and nearly temperature-independent surface conductivity. By performing the experiments from 2 \textendash ~20 K, we extracted the bulk conductivity, which includes both the bulk-dominated and surface-dominated regimes. We note that the numerically found geometric prefactors were used to extract the bulk conductivity, and a noticeable discontinuity between the results from the bulk-dominated regime and the surface-dominated regime existed, suggesting there is likely a discrepancy in the dimensions between the ideally suggested transport geometry and the realized samples. However, the activation energy, or the slope of the bulk conductivity vs. $1/T$, are consistent in the two regimes, suggesting that they can be patched together by adjusting the prefactors. We recommend adjusting the prefactor corresponding for the inverted resistance measurement. Using these methods, we have found the bulk conductivity of SmB$_{6}$, extending about two extra orders of magnitude compared to the conventional resistance measurement. 

We expect our newly proposed method of transport can be used in a broad range of materials beyond SmB$_{6}$. For these new materials, we hope that the community does not rely on the conventional transport methods such as residual-resistance ratios that can be problematic in the presence of two-channels, and instead, employ our transport methods. 

\begin{acknowledgments}
The authors thank Alexa Rakoski, Juniar Lucien, Meredith Henstridge, and Steven Wolgast for helpful discussions and comments on the manuscript. Funding for this work was provided by NSF Grants No. DMR-1441965, No. DMR-1643145, PHY1402971, and also from the Alfred P. Sloan Foundation. The cryogenic measurements were performed using a PPMS, which was acquired through an National Science Foundation MRI Award No. 1428226.
\end{acknowledgments}

\appendix
\section{\label{Appedix:AnalCorbino}Analytical Approach to Two-Terminal Corbino Disk with Infinite Thickness}

In this appendix, we take a deeper look at the perturbative approach of the scalable resistance that was introduced in Sec.~\ref{sec:TransportGeometry}. We will consider the boundary conditions that make the electrostatics unique when there are two conductive channels (bulk and surface). Using these findings, we will calculate the geometric coefficient ($C_1$) in the surface-dominated regime in an analytic fashion in the case of a standard two-terminal Corbino disk with infinite thickness. 

\subsection{Continuity Equation and Boundary Conditions in the Presence of Two Channels}

Consider a conductor with isotropic bulk conductivity, $\sigma_b$, and surface conductivity, $\sigma_s$. Within the linear response, the DC transport behavior can be characterized by the electric potential, $V(\vec{r})$, which is a scalar function of real space coordinate, $\vec{r}$. In the bulk of the material, $V(\vec{r})$ satisfies the Laplace equation:

\begin{equation}
\nabla^2 V(\vec{r}) = 0,
\label{Eq:AppendixLaplaceEq}
\end{equation}
which implies that the bulk current density is proportional to the electric field in the following manner: $\vec{J} = \sigma_{b}\vec{E} = -\sigma_{b}\nabla V$ and $\nabla\cdot\vec{J} = -\sigma_{b} \nabla^2 V =0$.
 
 On the surface, at a surface area covered with a highly conductive metal, the electric potential in this surface area is a constant:
\begin{equation}
V(\vec{r}) = \mathrm{constant}.
\end{equation}

At the surface area that is exposed, we have the following boundary conditions from the continuity equation:
\begin{equation}
\nabla_{s}\cdot\vec{j}_s = J_{n},
\label{Eq:AppendixBC}
\end{equation}
where the left-hand side is the 2D divergence ($\nabla_{s}\cdot$) of the surface current density $\vec{j}_s$, and on the right-hand side, $J_n$  is the current density flowing from the bulk to the surface in the normal direction of the surface. Note that $J_n$ has units of A/m$^2$, whereas $\vec{j}_s$ has units of A/m. Alternatively, Eq.~(\ref{Eq:AppendixBC}) can be expressed as: 
\begin{equation}
\nabla_{\parallel}^2 V(\vec{r}) = \frac{\sigma_b}{\sigma_s}\frac{\partial V}{\partial n},
\label{Eq:AppendixBCInfiniteThick}
\end{equation}
where $\nabla_{\parallel}^2$ is the 2D Laplacian on the surface and $\frac{\partial V}{\partial n}$ is the first order derivative along the normal direction of the surface. Here, $\sigma_b$ has units of 1/($\mathrm{\Omega}\cdot$m), whereas $\sigma_s$ has units of 1/$\mathrm{\Omega}$. 

For any transport geometry, the boundary conditions discussed above are uniquely determined by the geometric shape of the sample, the locations and the sizes of the leads, and the external electric potentials applied to each lead. With these boundary conditions, Eq.~(\ref{Eq:AppendixLaplaceEq}), has a unique solution. In other words, the spatial distribution of the current is uniquely determined in the bulk and on the surfaces, and therefore directly dictates all measured transport coefficients.  

\subsection{Two-Terminal Corbino Disk Geometry with Infinite Thickness}
 Here we consider the boundary conditions of a two-terminal Corbino disk, as shown in Fig.~(\ref{Fig:StandTwoTermCorbino}), in the limit where the thickness of the sample is infinite. 

 As discussed previously, the potential, $V$, follows Eq.~(\ref{Eq:AppendixLaplaceEq}) in the bulk. We choose cylindrical coordinates, $V(r,\phi,z)$. The coordinates are defined such that the top surface that has a Corbino disk is located at $z$ = 0, and the bulk occupies the entire $z<0$ area. For the boundary conditions, $V=0$ when $z=0$ and $r>r_{\mathrm{out}}$ (location of ground), and follow $V$ = constant for $r>r_{\mathrm{in}}$ (location of current source). Within the annulus region, $r_{\mathrm{in}}< r < r_{\mathrm{out}}$, the following boundary condition holds: 
\begin{equation}
\nabla_{\parallel}^2 V(\vec{r}) = \frac{\sigma_b}{\sigma_s}\frac{\partial V}{\partial z}.
\end{equation}
In addition, $V$ never diverges even in the limit of $r \rightarrow \infty$ and $z \rightarrow - \infty$.

\subsection{Perturbation Approach to Electric Potential}

The partial differential equation and the boundary conditions above are governed by the ratio between the bulk and the surface conductivity. We define a parameter, $\lambda=\sigma_b/\sigma_s$, and expand $V$ in power series: 
\begin{equation}
V = \sum_{i=0}^{+\infty}\lambda^i \varphi_i,
\label{Eq:AppendixPotSeries}
\end{equation}
where $\varphi_i$'s are expansion coefficients. Note that $\lambda$ is not dimensionless, whereas $x$ in Sec.~\ref{sec:Formalism} was defined so that it is dimensionless for convenience. 

 With Eq.~(\ref{Eq:AppendixLaplaceEq}), we find the following conditions of each $\varphi$ that must hold for all $i$:
\begin{equation}
\nabla^2 \varphi_i =0.
\label{Eq:laplaceforcoefficient}
\end{equation}
In the annular region on the surface, $r_{\mathrm{in}}<r<r_{\mathrm{out}}$ and $z=0$, $\nabla_{\parallel}^2 \varphi_{0} = 0$ when $i=0$. Also $\varphi_{i}$ = constant for $r<r_{\mathrm{in}}$, and $\varphi_i = 0$ for $r>r_{\mathrm{out}}$ on the surface ($z=0$). Therefore, we have:
\begin{equation}
\varphi_{0} (r,z=0)=\left\{
                \begin{array}{ll}
                  \frac{1}{2\pi\sigma_{s}} \ln (\frac{r_{\mathrm{in}}}{r_{\mathrm{out}}}) & ~~\mathrm{when}~r<r_{\mathrm{in}}\\
                  \frac{1}{2\pi\sigma_{s}} \ln (\frac{r}{r_{\mathrm{out}}}) &~~ \mathrm{when}~r_{\mathrm{in}}<r<r_{\mathrm{out}}\\
                  0 & ~~\mathrm{when} ~r>r_{\mathrm{out}}
                \end{array}
              \right.
\end{equation}
In the bulk, the solution of Eq.~(\ref{Eq:laplaceforcoefficient}) when $i=0$ that is continuous to the surface conditions above is:
\begin{equation}
\varphi_0 (r,z) = \frac{1}{2\pi\sigma_s}\int dk \frac{J_{0}(kr_{\mathrm{in}}) - J_{0}(kr_{\mathrm{out}})}{k}J_{0}(kr)e^{kz},
\label{Eq:FirstOrderSolution}
\end{equation}
where $J_{0}$ is a Bessel function of the first kind. Next, we find the relations for $i>0$. Plugging Eq.~(\ref{Eq:AppendixPotSeries}) into Eq.~(\ref{Eq:AppendixBCInfiniteThick}) and comparing the powers of $\lambda$, we have the following recursion relation:
\begin{equation}
\nabla_{\parallel}^2\varphi_{i} = \frac{\partial \varphi_{i-1}}{\partial z}
\end{equation}
when $i>1$. On the surface, $z=0$, and in the annular region, $r_{\mathrm{in}}<r<r_{\mathrm{out}}$, is:
\begin{equation}
\begin{split}
\varphi_{i} (r,z=0) = & -\ln (\frac{r_{\mathrm{out}}}{r})\int_{r_{\mathrm{in}}}^{r} dr_{0}~r_{0}\frac{\partial}{\partial z}\varphi_{i-1}(r_0, z)|_{z=0} \\
& -\int_{r}^{r_{\mathrm{out}}} dr_{0}~r_{0}\frac{\partial}{\partial z}\varphi_{i-1}(r_0, z)|_{z=0}\ln (\frac{r_{\mathrm{out}}}{r_0}).
\end{split}
\label{Eq:AppendixSurfaceRecursion}
\end{equation}
Thus, the boundary conditions for $\varphi_i$ can be found from $\varphi_{i-1}$. With the boundary conditions, the bulk solution of $\varphi_i$ can be obtained by:
\begin{equation}
\varphi_{i}(r,z) = \int_{0}^{+\infty}dk~A_{i}(k) J_{0}(kr)e^{kz},
\end{equation}
where
\begin{equation}
A_{i}(k) = k \int_{0}^{+\infty}dr~ rJ_{0}(kr) \varphi_{i}(r,z=0).
\end{equation}
Then, from $\varphi_{i}(r,z)$, we can use the same method shown above to obtain the boundary conditions for $\varphi_{i+1}(r,z)$, and the iteration can generate solutions to higher orders. 

\subsection{First Order Solution of Resistance}

From Eq.~(\ref{Eq:FirstOrderSolution}), we can compute the derivative of $\varphi_0$ along the $z$ direction:
\begin{equation}
\frac{\partial \varphi_{0}}{\partial z}|_{z=0}=\left\{
                \begin{array}{ll}
                  \frac{1}{2\pi\sigma_{s}} [\frac{2}{\pi r_{\mathrm{in}}} K (\frac{r^2}{r_{\mathrm{in}}^2}) - \frac{2}{\pi r_{\mathrm{out}}} K (\frac{r^2}{r_{\mathrm{out}}^2})] & ~~\mathrm{when}~r<r_{\mathrm{in}}\\
                  \frac{1}{2\pi\sigma_{s}} [\frac{2}{\pi r} K (\frac{r_{\mathrm{in}}^2}{r^2}) - \frac{2}{\pi r_{\mathrm{out}}} K (\frac{r^2}{r_{\mathrm{out}}^2})] & ~~\mathrm{when}~ r_{\mathrm{in}}<r<r_{\mathrm{out}}\\
                  \frac{1}{2\pi\sigma_{s}} [\frac{2}{\pi r} K (\frac{r_{\mathrm{in}}^2}{r^2}) - \frac{2}{\pi r} K (\frac{r_{\mathrm{out}}^2}{r^2})] & ~~\mathrm{when} ~r>r_{\mathrm{out}}
                \end{array}
              \right.,
\label{Eq:AppendixNormalFirstORderDeriv}
\end{equation}
where $K(x)$ is the complete elliptic integral of the first kind. Using Eq.~(\ref{Eq:AppendixNormalFirstORderDeriv}) with Eq.~(\ref{Eq:AppendixSurfaceRecursion}), the first-order correction of the potential difference between the two leads (source and ground) can be found as:
\begin{equation}
\Delta\varphi_{1}=\varphi_{1}(r_{\mathrm{in}},z=0) - \varphi_{2}(r_{\mathrm{in}},z=0) = - \frac{1}{\pi^2\sigma_s}w(r_{\mathrm{in}},r_{\mathrm{out}})
\end{equation}
where
\begin{equation}
\begin{split}
w(r_{\mathrm{in}},r_{\mathrm{out}}) = & \frac{1}{16R_{out}}[\pi r_{\mathrm{out}}^2{}_4F_{3}(\frac{1}{2},1,1,\frac{3}{2};2,2,2;1)\\
& -\pi r_{\mathrm{in}}^2 ( {}_4F_{3}(1,1,\frac{3}{2},\frac{3}{2};2,2,2;\frac{r_{\mathrm{in}}^2}{r_{\mathrm{out}}^2}) + {}_4F_{3}(\frac{1}{2},1,1,\frac{3}{2};2,2,2;\frac{r_{\mathrm{in}}^2}{r_{\mathrm{out}}^2}))\\
& -16r_{\mathrm{out}}(2GR_{\mathrm{out}} - 3r_{\mathrm{out}}E(\frac{r_{\mathrm{in}}^2}{r_{out}^2}) + 2r_{\mathrm{in}} + r_{\mathrm{out}})\\ 
& + 32 (r_{\mathrm{in}}^2 - r_{\mathrm{out}}^2) K(\frac{r_{\mathrm{in}}^2}{r_{\mathrm{out}}^2})\\
& + 16\ln(\frac{r_{\mathrm{out}}}{r_{\mathrm{in}}})( (r_{\mathrm{in}}^2 - r_{\mathrm{out}}^2)K(\frac{r_{\mathrm{in}}^2}{r_{\mathrm{out}}^2}) + r_{\mathrm{out}}^2 E(\frac{r_{\mathrm{in}}^2}{r_{\mathrm{out}}^2}) - r_{\mathrm{in}}r_{\mathrm{out}})\\
& + 16 \pi r_{\mathrm{out}}^2 \ln(2) ].\\
\end{split}
\end{equation}
Here, ${}_{p}F_{q}(a;b;z)$ is the generalized hypergeometric function, $K(x)$ is a complete elliptic integral of the first kind, and $G$ is the Catalan constant. In addition to the electrical potential, there is also a correction to the current by the perturbation expansion. The corrections to the current to first order is:
\begin{equation}
I_{1} = \sigma_{b}\int_{0}^{r_{\mathrm{in}}}2\pi r dr \frac{\partial \varphi_{0}}{\partial z}|_{z=0}.
\end{equation}
Using Eq.~(\ref{Eq:AppendixNormalFirstORderDeriv}), we find:
\begin{equation}
I_{1} = \frac{2I}{\pi} \frac{\sigma_b}{\sigma_s}\nu(r_{\mathrm{in}},r_{\mathrm{out}}),
\end{equation}
where 
\begin{equation}
\nu(r_{\mathrm{in}},r_{\mathrm{out}}) = r_{\mathrm{in}} - \frac{r_{\mathrm{in}}^2 K(\frac{r_{in}^2}{r_{out}^2})}{r_{\mathrm{out}}} + r_{\mathrm{out}} K(\frac{r_{in}^2}{r_{out}^2}) - r_{\mathrm{out}} E(\frac{r_{in}^2}{r_{out}^2}).
\end{equation}
Thus, the resistance including the first order correction is:
\begin{equation}
\begin{split}
R &=  \frac{V}{I} \approx \frac{\Delta \varphi_{0} + \frac{\sigma_b}{\sigma_s} \Delta \varphi_{1}}{I+I_{1}}\\
&= \frac{\ln(r_{\mathrm{out}}/r_{\mathrm{in}})}{2\pi\sigma_{s}} \frac{ 1 - \frac{2}{\pi}\frac{1}{\ln(r_{\mathrm{out}}/r_{\mathrm{in}})}\frac{\sigma_b}{\sigma_s}w(r_{\mathrm{in}}, r_{\mathrm{out}})     }{ 1 + \frac{2}{\pi} \frac{\sigma_b}{\sigma_s}\nu(r_{\mathrm{in}}, r_{\mathrm{out}})      }   \\
\end{split}
\end{equation}
To the first order, this is equivalent to:
\begin{equation}
R \approx \frac{\ln(r_{\mathrm{out}}/r_{\mathrm{in}})}{2\pi\sigma_{s}} \frac{1}{ 1 + \frac{\sigma_b}{\sigma_s}\gamma(r_{\mathrm{in}}, r_{\mathrm{out}})},
\end{equation}
or alternatively, defining $\alpha=r_{\mathrm{out}}/r_{\mathrm{in}}$, $\gamma$ can be expressed as:
\begin{equation}
\begin{split}
\gamma (\alpha) = & - \frac{r_{\mathrm{in}}}{\ln(\alpha)}[ \frac{4}{\pi} + \alpha (\frac{2}{\pi} - 2 \ln2) + \frac{4}{\pi}G\alpha\\
& -\frac{6}{\pi}\alpha E(\alpha^{-2}) + \frac{4}{\pi}(\alpha - 1/\alpha)K(\alpha^{-2})\\
& - \frac{\alpha}{8}{}_{4}F_{3}(\frac{1}{2},1,1,\frac{3}{2};2,2,2;1) + \frac{1}{8\alpha}{}_{4}F_{3}(\frac{1}{2},1,1,\frac{3}{2};2,2,2;\alpha^{-2})].\\
\end{split}
\end{equation}

When $\alpha =2$, $\gamma(2)/r_{\mathrm{in}} = 0.626792$, which agrees with the numerical results in the infinite thickness limit that is shown in Fig.~(\ref{Fig:StandTwoTermCorbinoResult}).

\section{Generalization to Anisotropic Bulk Conductivity}

The work in this paper so far has been done under the assumption that the bulk conductivity is isotropic. We know that many topological insulators are in fact anisotropic in such a way that the material has a different conductivity in one specific crystal direction. If we prepare a double-sided Corbino disk sample where the axis is aligned with that specific direction, then in principle, we can use the bulk conductivity measurement strategies by theoretically mapping the anisotropic sample to an isotropic sample with a different thickness. 

For a material with anisotropic bulk conductivity, we define the direction that is tangential to the surface as $r$, and normal to the surface as $z$. We also define $\sigma_{tn}$ as the bulk conductivity tangential to the surface, and $\sigma_{n}$ as the bulk conductivity normal to the surface. Then, the potential in the bulk is determined by 
 \begin{equation}
	\sigma_{tn}\frac{\partial^2V}{\partial r^2}+\sigma_{n}\frac{\partial^2V}{\partial z^2}=0,
	\label{Eq:AnisotropicLaplaceEq}
\end{equation} 
with the appropriate boundary conditions at the interface of bulk and surface. By redefining the normal direction, $z$, to $\zeta=\sqrt{\sigma_{tn}/\sigma_{n}}z$, we can re-write Eq.~(\ref{Eq:AnisotropicLaplaceEq}) as:
\begin{equation}
	\frac{\partial^2V}{\partial r^2}+\frac{\partial^2 V}{\partial\zeta^2}=0,
	\label{Eq:MapptedAnisotropicLaplaceEq}
\end{equation} 
Thus, we have mapped our problem into an isotropic bulk conductivity case. When we map the problem as above, the thickness of the sample, $t$, is also mapped to a different thickness, $\tau$, with the relation:  $\tau=\sqrt{\sigma_{tn}/\sigma_{n}}t$. 

$\sigma_{tn}$ can be found if we know the correct $C_{1}(\tau)$ and $C_{-1}(\tau)$ values. If $\tau$ is not found correctly, $\sigma_{tn}$ in the surface-dominated regime and the bulk-dominated regime will be mismatched. If we start with the bulk conductivity from $C_{1}(t)$ and $C_{-1}(t)$, which is mismatched initially, and change those values to $C_{1}(\tau)$ and $C_{-1}(\tau)$, where the bulk conductivity in two regimes are continuous, this corresponds to $\sigma_{tn}$. Finally, $\sigma_{tn}$ and $\tau$ can be used to find the bulk conductivity in the normal direction: 
\begin{equation}
	\sigma_{tn}=\sigma_{n}(\frac{t}{\tau})^2.
	\label{Eq:MapptedAnisotropicConductivity}
\end{equation} 
However, we should warn the reader that this strategy for anisotropic conductivities can only be useful for samples that are prepared with extremely high precision. As we have learned from our transport measurements on SmB$_{6}$ samples, a mismatch between the bulk conductivity extracted from bulk- and surface-dominated regimes using different 4-terminal resistance measurements can easily occur because of small imperfections in the transport geometries.

\bibliography{Inverted}

\end{document}